\documentclass[aps,prc,reprint,superscriptaddress,floatfix,showpacs]{revtex4-1}
\usepackage{graphicx}
\usepackage{amssymb}
\usepackage{dcolumn}

\begin{document}

\title{Magnetism of PdNi alloys near the critical concentration for ferromagnetism}

\author{G.~M. Kalvius}\email[To whom correspondence should be addressed. Email address: ]{kalvius@ph.tum.de.}
\affiliation{Physics Department, Technical University Munich, 85747 Garching, Germany}
\author{O. Hartmann}
\author{R. W\"appling}
\affiliation{Department of Physics and Astronomy, Uppsala University, 75120 Uppsala, Sweden}
\author{A. G\"unther}
\author{A. Krimmel}
\author{A. Loidl}
\affiliation{Experimental Physics V, Center for Electronic Correlations and Magnetism, University of Augsburg, 86159 Augsburg, Germany}
\author{D.~E. MacLaughlin}
\affiliation{Department of Physics and Astronomy, University of California, Riverside, California 92521, USA}
\author{O. O. Bernal}
\affiliation{Department of Physics and Astronomy, California State University, Los Angeles, California 90032, USA}
\author{G.~J. Nieuwenhuys}
\altaffiliation{Deceased.}
\affiliation{Laboratory for Muon-Spin Spectroscopy, Paul Scherrer Institute, CH-5232 Villigen-PSI, Switzerland}
\author{M.~C. Aronson}
\affiliation{Condensed Matter Physics and Materials Science Department, Brookhaven National Laboratory, Upton, New York 11973, USA}
\affiliation{Department of Physics and Astronomy, Stony Brook University, Stony Brook, New York 11794, USA}
\author{R.~P. Dickey}
\author{M.~B. Maple}
\affiliation{Department of Physics, University of California, San Diego, La Jolla, California 92093, USA}
\author{A. Amato}
\author{C. Baines}
\affiliation{Laboratory for Muon-Spin Spectroscopy, Paul Scherrer Institute, CH-5232 Villigen-PSI, Switzerland}

\date{\today}

\begin{abstract}

We report results of a muon spin rotation and relaxation ($\mu$SR) study of dilute Pd$_{1-x}$Ni$_x$ alloys, with emphasis on Ni concentrations~$x = 0.0243$ and 0.025. These are close to the critical value~$x_\mathrm{cr}$ for the onset of ferromagnetic long-range order (LRO), which is a candidate for a quantum critical point. Additional control data were taken for pure nonmagnetic Pd, and for an alloy where ferromagnetism is well established ($x = 0.05$). The 2.43 and 2.5~at.\,\%~Ni alloys exhibit similar $\mu$SR properties. Both samples are fully magnetic, with average muon local fields $\langle B^\mathrm{loc}(T{=}0)\rangle = 2.0$ and 3.8~mT and Curie temperatures~$T_C = 1.0$ and 2.03~K for 2.43 and 2.5~at.\,\%~Ni, respectively. The temperature dependence of $\langle B^\mathrm{loc}\rangle$ suggests ordering of Ni spin clusters rather than isolated spins. Just above $T_C$, the temperature where LRO vanishes, a two-phase region is found with coexisting separate volume fractions of quasistatic short-range order (SRO) and paramagnetism. The SRO fraction decreases to zero with increasing temperature a few kelvin above $T_C$. This mixture of SRO and paramagnetism is consistent with the notion of an inhomogeneous alloy with Ni clustering. The measured values of $T_C$ extrapolate to $x_\mathrm{cr} = 0.0236 \pm 0.0027$. The dynamic muon spin relaxation in the vicinity of $T_C$ differs for the two samples: a relaxation-rate maximum at $T_C$ is observed for $x = 0.0243$, reminiscent of critical slowing down, whereas for $x = 0.025$ no dynamic relaxation is observed within the $\mu$SR time window. The data suggest a mean-field-like transition in this alloy.
\end{abstract}


\maketitle

\section{\label{intr} Introduction}

Palladium is a highly exchange enhanced paramagnet close to a ferromagnetic instability. A ferromagnetic state with long-range order (LRO) can be achieved by alloying with $3d$ elements, in particular Ni. Consequently, most studies of the PdNi system concentrated on the mechanism of magnetic moment formation (see, e.g., Refs.~\cite{ARS70} and ~\cite{HaZu72}). Of particular interest is the critical Ni concentration $x_\mathrm{cr}$ for the onset of ferromagnetism. In an early study~\cite{CrSc65} a value of $\sim$2.6~at.\,\%~Ni was reported. Later work, based mainly on low field magnetization data, used linear extrapolation and found critical compositions of 2.3~at.\,\% Ni and 2.5~at.\,\%~Ni (Refs.~\cite{MTC74} and ~\cite{CKG81}, respectively). Quasielastic neutron scattering in Pd(1~at.\,\%~Ni)~\cite{BRLM83} revealed short-lived spin fluctuations (paramagnons) and a $Q$-dependence of the magnetic scattering that essentially follows the Ni magnetic form factor. 

Pd$_{1-x}$Ni$_x$ in the regime close to $x_\mathrm{cr}$ is a prime candidate to exhibit a quantum critical point (QCP), provided structural and magnetic disorder are small. A QCP is characterized by the interplay between thermal and quantum fluctuations when an energy characteristic for the order parameter fluctuation exceeds the thermal energy. Microscopically, the peculiar properties in the quantum critical regime, such as fractional power laws in thermodynamic quantities, can be ascribed to unconventional elementary excitations. The standard (Fermi liquid) model of metals fails in this realm. For example, the specific-heat coefficient~$\gamma = C/T$ often diverges as $T \to 0$, as opposed to the asymptotic Fermi-liquid result~$C/T = \mathrm{const.}$ The temperature-dependent part of the electrical resistivity varies proportionally to $T^m$ with $m$ less than the Fermi-liquid value of 2. The presence of non-Fermi liquid (NFL) behavior is considered a hallmark of a magnetic QCP in metals. T.~Vojta~\cite{Vojt10} has reviewed rare-region effects at quantum phase transitions in disordered metals.

An investigation of Pd$_{1-x}$Ni$_x$ in the low-Ni-concentration limit using electrical resistivity, specific heat and magnetic susceptibility measurements~\cite{NBKM99} indicated NFL behavior, and suggested the presence of a QCP at $x_\mathrm{cr} = 0.026 \pm 0.002)$. This value was established by fitting the Curie temperature~$T_C(x)$ to the expected behavior $T_C \propto (x - x_\mathrm{cr})^{3/4}$. The canonical behavior expected for a pure ferromagnet at a QCP was observed for the electrical resistivity ($\rho \propto T^{5/3}$), the specific heat ($C/T \propto -\ln T$), and the magnetic susceptibility ($\chi = \chi_0 - \chi_1T^{3/4}$)~\cite{Lonz97,*Stew01}. In contrast, similar studies with respect to a ferromagnetic QCP under compositional tuning performed for Pd doped with Mn~\cite{HMW82} or Fe~\cite{Nieu75,PBKM84} revealed spin-glass behavior with substantial spin disorder. Dilute \textit{Pd}Fe alloys exhibit giant-moment behavior giving rise to inhomogeneous magnetization, and are viewed as magnetic cluster systems~\cite{KSP95}.

Only one muon spin rotation/relaxation ($\mu$SR) investigation of PdNi alloys has been reported~\cite{RCKS94}: a study of alloys with 3.3 and 5.8~at.\,\%~Ni, well within the established ferromagnetic regime. The principal subject of that study was again the question of magnetic moment formation in an impurity system. The characteristic behavior of critical spin fluctuations was observed when approaching $T_C$ (20.5 and 90~K, respectively) from above in weak transverse fields, but no muon relaxation at all was observed in zero-field data above or below $T_C$. The authors attributed this unusual result to rapid muon diffusion, so that in zero field muons average over magnetic domains with random magnetization orientations. In pure Pd our results are consistent with rapid muon diffusion as observed earlier~\cite{HHWH92}, but in PdNi alloys we find the expected behavior for a stationary muon in an ordered magnetic material (Sec.~\ref{sec:res}). Muons are localized by very light doping ($\sim$100~ppm) in bcc V and Nb and in fcc Al, so that Ni-induced localization in Pd is not surprising. An overview of muon diffusion in solids is given in Ref.~\cite{Karl95}.

The aim of the present $\mu$SR investigation was to gain information on atomic-scale magnetic properties of PdNi alloys near a possible QCP\@. $\mu$SR is a useful technique in part because it samples the magnetic field~$\mathbf{B}^\mathrm{loc}$ at the muon site due to its local magnetic environment (within a few lattice sites), and is therefore not dependent on coherent diffraction from a periodic structure. Thus $\mu$SR can give unique information on magnetism even in disordered systems with only short-range order (SRO) (e.g., spin glasses). By the same token, $\mu$SR normally cannot determine magnetic correlation lengths directly~\cite{[{See, however, }] Noak99}. We use the terms LRO and SRO qualitatively, based on indirect information: LRO denotes the situation where magnetic moments are correlated over long enough distances to yield a well-defined magnitude~$B^\mathrm{loc}$, whereas for SRO the local disorder leads to a spread in $B^\mathrm{loc}$ that dominates the response of the muon~\cite{Noak99}. The two cases can be distinguished by the shapes of their spectra as discussed in Sec.~\ref{p2p4n}.

Other useful features of $\mu$SR include a unique spectral window (the megahertz frequency range) for slow spin dynamics, and the fact that in cases of coexisting magnetic phases one can derive the volume fraction of each phase directly (i.e., without corrections such as the Debye-Waller factor or saturation effects) from the strength of its $\mu$SR signal. 

The major results of this work come from two dilute alloys with compositions close to $x_\mathrm{cr}$, which were prepared independently and investigated in independent $\mu$SR experiments by independent subgroups of the present authors. The structure of the paper is as follows. After describing the samples and their preparation and reviewing the $\mu$SR technique in Sec.~\ref{sec:exp}, we briefly discuss results for pure Pd metal and a Pd(5~at.\,\%~Ni) alloy in Secs.~\ref{sec:ppd} and \ref{sec:p5n}, respectively. These were studied in order to get a feeling for $\mu$SR in definitely nonmagnetic and well-developed magnetic alloys in the PdNi series. 
In Sec.~\ref{p2p4n} we discuss the data analysis procedures and the basic experimental findings for the case of a 2.43~at.\,\%~Ni alloy, for which the experimental data are somewhat more complete. In Sec.~\ref{sec:p2p5n} we present the results for a 2.5~at.\,\%~Ni alloy, for which the bulk magnetic, transport, and thermal properties have been well characterized~\cite{NBKM99}. We then compare the properties of the two closely-related materials. The results with respect to the local magnetic properties and moment dynamics are discussed and summarized in Sec.~\ref{sum}.

\section{\label{sec:exp} Experiment}

Our high-purity palladium sample (99.99+\% purity with 3ppm Fe) consists of a stack of thin foils that had been used previously for studies of muon diffusion~\cite{HHWH92}. The 5~at.\,\%~Ni and 2.5~at.\,\%~Ni samples were $\sim$1~cm-diameter buttons from the batch used previously for bulk property measurements~\cite{NBKM99}. They were prepared from high purity (5N) starting materials by an argon-arc technique. The samples were remelted several times and finally annealed for 5 days at 1000$^\circ$C\@. X-ray diffraction and microprobe analysis confirmed the absence of spurious phases. The Pd(2.43~at.\,\%~Ni) sample was prepared by remelting the constituents many times, followed by a 48 hour anneal at 1200$^\circ$C and a subsequent quench in ice water. The compositional homogeneity of the arc-melted pellets was investigated using an electron microprobe. We found that the Ni concentration was slightly elevated near the surface of the pellets, but uniform within the resolution limit of the microprobe ($\sim$5\%) for most of the pellet interior.

For details of the time-differential $\mu$SR technique we refer the reader to monographs and review articles~\cite{Sche85,Karl95,LKC99,KNH01,YaDdR11}. Briefly, spin-polarized (usually positive) muons are stopped in the sample (usually at an interstitial site), and decay via the reaction~$\mu^+ \to e^+ + \nu_e + \overline{\nu}_\mu$. The decay positrons are emitted asymmetrically, preferentially in the direction of the muon spin at the time of decay. The goal of the experiment is to determine the time evolution of the muon spin polarization function~$G(t)$, which contains information on the magnitude, static distribution, and fluctuations of the local magnetic field~$\mathbf{B}^\mathrm{loc}$ due to neighboring currents and/or magnetic moments in the sample. The quantity measured is the time dependence of the positron count-rate asymmetry
\begin{equation} \label{asy}
A(t) = A(0)G(t) \,,
\end{equation}
where $A(0)$ is the initial asymmetry (typically 0.2--0.25). The total field~$\mathbf{B}_\mu$ at the muon site is given by $\mathbf{B}_\mu = \mathbf{B}^\mathrm{loc} + \mathbf{B}^\mathrm{int}$, where $\mathbf{B}^\mathrm{int}$ is the internal field due to the external field~$\mathbf{B}^\mathrm{ext}$; $\mathbf{B}^\mathrm{int} \ne \mathbf{B}^\mathrm{ext}$ if macroscopic effects such as the demagnetizing field are significant. Magnetic properties of the system are inferred from the effect of $\mathbf{B}^\mathrm{loc}$ on the muon spin. 

$\mu$SR data were taken between 0.02 and 300~K in weak transverse fields, longitudinal fields, and zero field (TF-$\mu$SR, LF-$\mu$SR, and ZF-$\mu$SR, respectively) at the General Purpose Spectrometer (GPS) (2--300~K) and Low Temperature Facility (LTF) (0.02--2~K) of the Paul Scherrer Institute (PSI), Villigen, Switzerland. Weak transverse fields were generated by auxiliary Helmholtz coils, and the longitudinal field at the LTF spectrometer was supplied by a superconducting split-pair magnet. At the GPS veto counters eliminate the background signal from muons that do not stop in the sample, a feature that is not available at the LTF.

The incident muon beam, with muon spins parallel to the beam direction, is passed through a ``separator'' with crossed electric and magnetic fields before stopping in the sample, in order to remove unwanted particles (mostly positrons) from the beam. The separator magnetic field also rotates the muon spin, so that for LF-$\mu$SR experiments the muon spin is unavoidably tilted a few degrees away from the external longitudinal field~$\mathbf{B}_L^\mathrm{ext}$.\footnote{The main magnetic field is applied parallel to the muon beam, since deviation of the beam is prohibitive for perpendicular fields greater than a few tens of millitesla. For TF-$\mu$SR in the main field, the separator fields are increased in order to rotate the muon spin $90^\circ$.} This slight tilt is useful, as it provides a small but measurable precession signal in the spectrometer ``side'' counters that gives the magnitude~$B_L^\mathrm{int}$ of the internal field. It is particularly useful for studies of ``soft'' ferromagnets in low $B_L^\mathrm{ext}$, where the high permeability and consequent large demagnetization field can reduce or even cancel $\mathbf{B}_L^\mathrm{int}$.\footnote{See, e.g., Ref.~\protect~\cite{YaDdR11}, Chap.~5}

\section{\label{sec:res} Results}

\subsection{\label{sec:ppd} Pure Pd}

Data were taken between 0.02~K and 2~K using the LTF spectrometer. Zero-field data gave a relaxation rate~$0.008~\mu\mathrm{s}^{-1}$ at all temperatures measured, consistent with previous results~\cite{HHWH92}. This is essentially the minimum value that can be measured reliably due to the finite muon lifetime, and is considerably smaller than the value ${\sim}0.03~\mu\mathrm{s}^{-1}$ expected from $^{105}$Pd nuclear moments~\cite{HHWH92}. TF-$\mu$SR data in a field of 7~mT at 0.02~K, however, gave a markedly larger rate of $0.044(2)\ \mu\mathrm{s}^{-1}$. The likely explanation is a slight inhomogeneity of the external transverse field. Our results confirm the previous conclusions~\cite{HHWH92} that pure Pd is nonmagnetic, and that the muon diffuses rapidly in it.

\subsection{\label{sec:p5n} Pd(5~at.\,\%~Ni)}

\begin{figure}[ht]
 \includegraphics[clip=,width=3in]{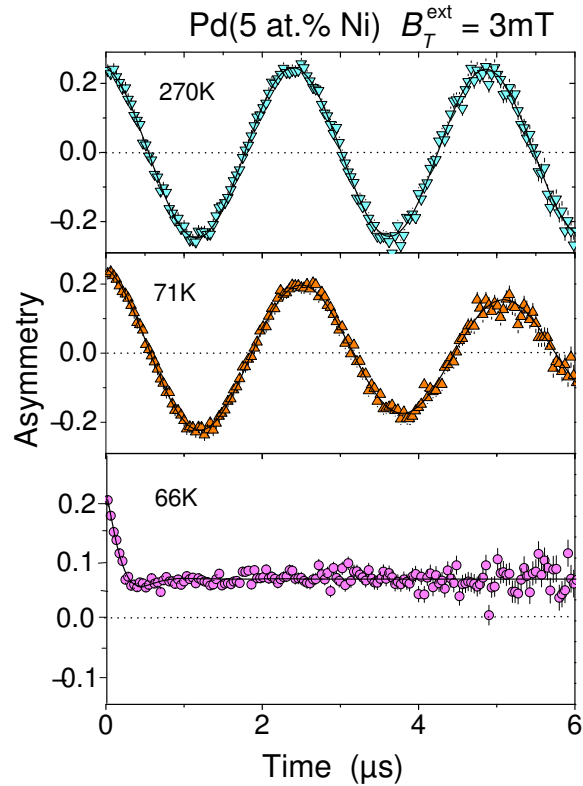}
 \caption{\label{p5n-tf} (Color online) TF-$\mu$SR asymmetry data for Pd(5~at.\,\%~Ni) at various temperatures in a transverse external field~$B_T^\mathrm{ext} = 3$~mT\@. Solid curves: fits (see text).}
\end{figure}

\begin{figure*}[ht]
 \includegraphics[clip=,width=5.5in]{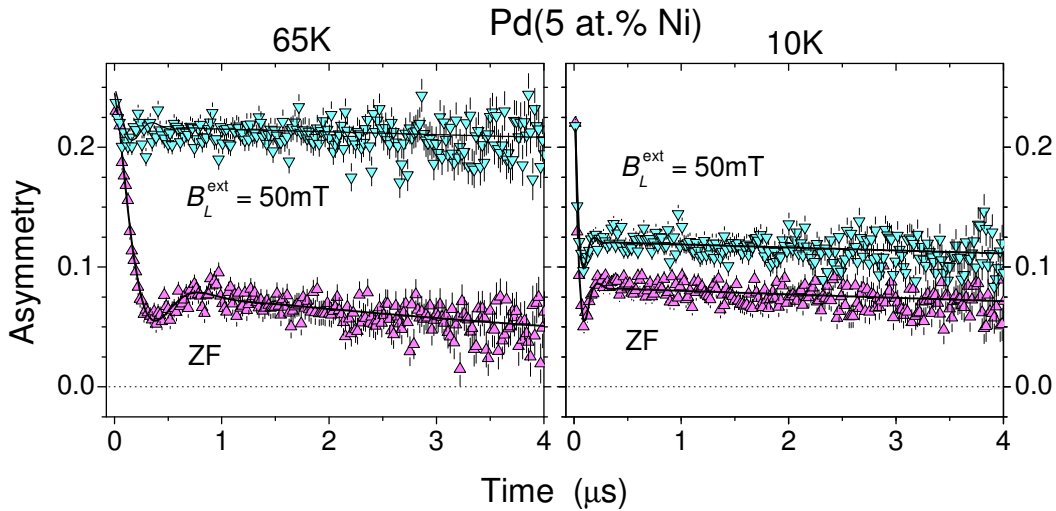}
 \caption{\label{fig:p5n-zf} (Color online) ZF- and LF-$\mu$SR asymmetry data for Pd(5~at.\,\%~Ni) in the ferromagnetic phase. Solid curves: fits of the OMAG function [Eq.~(\protect\ref{eq:omag})] to the data.} 
\end{figure*}

A concentration of 5~at.\,\% Ni in Pd is well above the critical value for the onset of ferromagnetic order. Data were taken between 10 and 270~K using the GPS spectrometer. TF-$\mu$SR asymmetry data in a transverse external field~$B_T^\mathrm{ext} = 3$~mT are shown in Fig.~\ref{p5n-tf}. 
The asymmetry data at 270 and 71~K exhibit the weakly damped muon spin rotation in $B_T^\mathrm{ext}$ characteristic of a paramagnetic sample. In contrast, the signal at 66~K resembles those observed at low temperatures in zero field (Fig.~\ref{fig:p5n-zf}). This is expected if the temperature is below $T_C$, so that for external fields less than the saturation value the high permeability results in a demagnetizing field that cancels the external field (cf.\ the discussion in Sec.~\ref{sec:exp}). Thus the TF data show that $66~\mathrm{K} < T_C < 71$~K, compared with 62.7~K obtained from the ac susceptibility~\cite{WKW90}. 

Figure~\ref{fig:p5n-zf} 
displays ZF- and LF-$\mu$SR ($B_L^\mathrm{ext} = 50$~mT) asymmetry data in the ferromagnetic state. Separate rapidly- and slowly-relaxing ZF signals are observed, the amplitudes of which are 2/3 and 1/3, respectively, of the total. This clearly indicates that the muons experience static or quasistatic (slowly varying on the muon time scale) local fields and hence do not diffuse.\footnote{Muon diffusion would result in ``motionally-narrowed'' relaxation characterized by a single exponential decay.} The reason for the two relaxation rates is as follows~\cite{KuTo67,HUIN79}: consider a sample for which $\mathbf{B}^\mathrm{loc}$ is static (i.e., due to frozen magnetic moments in the sample) and randomly oriented (e.g., in a powder). For a given muon site where the total field~$\mathbf{B}_\mu$ makes an angle~$\theta$ with the initial muon spin~$\mathbf{S}_\mu(0)$, the muon spin component~$S_\mu(0)\cos\theta$ parallel to $\mathbf{B}^\mathrm{loc}$ does not precess, while the perpendicular component~$S_\mu(0)\sin\theta$ precesses at the angular frequency~$\gamma_\mu B_\mu$. The projection of the non-precessing muon spin component onto the initial muon spin direction, which is also the axis of the LF-$\mu$SR counter system, is $S_\mu(0)\cos^2\theta$; this gives the ``longitudinal'' contribution to the LF-$\mu$SR signal. 

In a randomly-oriented polycrystalline sample in zero external field (ZF-$\mu$SR), the longitudinal contribution to the initial signal is proportional to the ``powder average''~$\langle\cos^2\theta\rangle = 1/3$. The remaining 2/3 of the initial signal is due to the transverse (precessing) muon spin components noted above. This transverse contribution is damped at the relaxation rate~$\lambda_T$, which is usually dominated by the static distribution of $B^\mathrm{loc}$. The relaxation rate~$\lambda_L$ of the longitudinal contribution is due solely to fluctuations of the electronic magnetic moments that generate a stochastic time dependence to $\mathbf{B}^\mathrm{loc}(t)$. Often $\lambda_L <\lambda_T$, in which case transverse relaxation dominates at early times and longitudinal relaxation at late times. Then the relaxation function has a characteristic ``2/3\,--\,1/3'' structure, as seen in the ZF-$\mu$SR data of Fig.~\ref{fig:p5n-zf}.

For both zero and nonzero external field the data of Fig.~\ref{fig:p5n-zf} were fit with the muon spin polarization function 
\begin{eqnarray}
 G_\mathrm{OMAG}(t) & = & (1-f_L) \exp(-\lambda_T t) \cos(\gamma_\mu\langle B_\mu\rangle t + \phi) \nonumber \\
& & \mbox{} + f_L \exp(-\lambda_L t)
 \label{eq:omag} 
 \end{eqnarray}
appropriate to an ordered magnetic (OMAG) material. Here $\lambda_T$ and $\lambda_L$ are the transverse and longitudinal relaxation rates discussed above, $\phi$ is a phase factor, $\gamma_\mu = 2\pi \times 135.5$~MHz/T is the muon gyromagnetic ratio, $\langle B_\mu\rangle$ is the spatial and temporal average of $B_\mu$, and $f_L$ is the fractional amplitude of the longitudinal component. In a randomly-oriented polycrystalline sample in zero external field we expect $f_L = 1/3$. 

Fits to the zero field asymmetry data for 65~K and 10~K using the OMAG function yield the parameter values given in Table~\ref{5pc-zfdat}.
\begin{table}[ht]
\caption{\label{5pc-zfdat} Transverse muon relaxation rate~$\lambda_T$, longitudinal relaxation rate~$\lambda_L$, and mean muon local field~$\langle B^\mathrm{loc}\rangle$ from fits to ZF-$\mu$SR asymmetry data below $T_C$ in Pd(5~at.\,\%~Ni).}
\begin{ruledtabular}
\begin{tabular}{cddd}
$T~(\mathrm{K})$ & \multicolumn{1}{c}{$\lambda_T~(\mu\mathrm{s}^{-1})$} & \multicolumn{1}{c}{$\lambda_L~(\mu\mathrm{s}^{-1})$} & \multicolumn{1}{c}{$\langle B^\mathrm{loc}\rangle~(\mathrm{mT})$} \\
\colrule
65 & 4.7 \pm 0.2 & 0.12\pm 0.02 & 8.5 \pm 0.5\\
10 & 16.5 \pm 1 & 0.04 \pm 0.01 & 30. \pm 3.\\
\end{tabular}
\end{ruledtabular}
\end{table}
The transverse damping rate is too large to allow the development of more than one oscillation period (Fig.~\ref{fig:p5n-zf}). At 10~K the half width at half maximum~$\Delta B^\mathrm{loc} = \lambda_T/\gamma_\mu$ of the local field distribution is 19~mT, which is 63\% of the mean local field~$\langle B^\mathrm{loc}\rangle = 30$~mT\@. At 65~K the ratio $\Delta B^\mathrm{loc}/\langle B^\mathrm{loc}\rangle$ is the same as at 10~K\@. Thus the ferromagnetic order is far from uniform, since strong local spin disorder is present independent of temperature below $T_C$.

The ZF-$\mu$SR asymmetry data in Fig.~\ref{fig:p5n-zf} clearly show that $\lambda_L$ depends on temperature; values are given in Table~\ref{5pc-zfdat}. This is in contrast to the temperature independence of $\lambda_L$ in the ferromagnetic regime reported in Ref.~\cite{RCKS94}. According to the present data, however, the ordered spin system shows the expected evolution toward the static limit when the temperature is reduced.

The fits to the LF-$\mu$SR data and their interpretation will be discussed in Sec.~\ref{sec:2p4lf}\@. We note here that in a longitudinal external field $\mathbf{B}_L$ the resultant total static local field~$\langle\mathbf{B}_\mu\rangle = \langle\mathbf{B}^\mathrm{loc}\rangle + \mathbf{B}_L$ is ``decoupled'' from $\langle\mathbf{B}^\mathrm{loc}\rangle$ and becomes nearly parallel to $\mathbf{S}_\mu(0)$ for $\mathbf{B}_L \gg \langle\mathbf{B}^\mathrm{loc}\rangle$. This increases the average~$\langle\cos^2\theta\rangle$ and hence the late-time longitudinal signal, as can be seen in Fig.~\ref{fig:p5n-zf}.

\subsection{\label{p2p4n} Pd(2.43~at.\,\%~Ni)} 

\subsubsection{Transverse-Field Data} \label{sec:2p4TF}

\begin{figure}[ht]
 \includegraphics[clip=,width=3in]{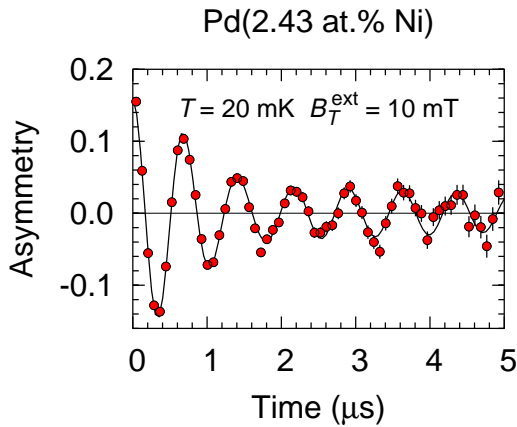}
 \caption{\label{fig:tf-2sig} (Color online) TF-$\mu$SR asymmetry data from Pd(2.43~at.\,\%~Ni), $T = 20$~mK, $B^\mathrm{ext}_T = 10$~mT\@. Curve: fit to the sum of sample and background signals.}
\end{figure}

In Fig.~\ref{fig:tf-2sig} 
TF-$\mu$SR asymmetry data for Pd(2.43~at.\,\%~Ni) taken at 20~mK in a transverse external field~$B_T^\mathrm{ext} = 10$~mT are shown, together with a fit to the sum of two terms: a damped oscillatory signal from the sample, and an undamped background signal from a fraction of muons that miss the sample and stop in the silver cold finger. Such a background signal is commonly observed in $\mu$SR experiments, unless a veto system is used as in the GPS\@. In the data of Fig.~\ref{fig:tf-2sig} from the LTF the background signal contributes about 12\% of the total asymmetry. The existence of an oscillatory signal from the sample shows that $B_T^\mathrm{ext}$ enters the bulk of the material. The field at the muon sites in the sample is $\sim$4\% higher than in the cold finger, reflecting the large susceptibility of the 2.43~at.\,\%~Ni alloy at low temperatures~\cite{Sche85a}. Fig.~\ref{tf-2sig} in Sec.~\ref{sec:p2p5n} shows the separation of analogous TF asymmetry data into a sample and a background signal for the case of Pd(2.5~at.\,\%~Ni).

In our ZF and LF data from the 2.43~at.\,\%~Ni alloy, discussed below, no ``silver background'' signal was evident. The least-squares fits, which included a fraction~$f_\mathrm{Ag}$ of silver background as a free parameter in the fitting, invariably yielded $f_\mathrm{Ag} = 0$ and were unacceptable if $f_\mathrm{Ag}$ were fixed at the value from the TF data. This is surprising, given the substantial background signal seen in TF-$\mu$SR (Fig.~\ref{fig:tf-2sig}). The sample was unusually large and thick, however, and in ZF- and LF-$\mu$SR evidently collected essentially all the muons. The difference in TF-$\mu$SR can be attributed to bending of the muon beam in the 10~mT transverse field, so that the beam spot moved by a few millimeters. This would have been enough to implant some of the muons in the cold finger.

\subsubsection{Zero- and Weak Longitudinal-Field Data, $T < T_C$} \label{sec:zfwftltc}

Data were taken in zero field at 20~mK and in a weak longitudinal field~$B_L^\mathrm{ext} = 1.1$~mT in the temperature range~0.02--1.4~K using the LTF\@. The weak field was necessitated by the presence of a small amount of trapped magnetic flux in the LTF superconducting magnet in later experiments; this produced a $\sim$100~$\mu$T remanent field of unknown orientation at the sample, capable of precessing the muon spin. The 1.1~mT longitudinal field ensured that the total resultant external field was sensibly parallel to the muon spin, thereby quenching such unwanted precession. Data were also taken in the GPS over the temperature range~1.56--4.5~K in essentially zero field ($\lesssim 10~\mu$T), since the resistive magnet of the GPS has no remanent field.

\begin{figure}[ht]
 \includegraphics[clip=,width=3in]{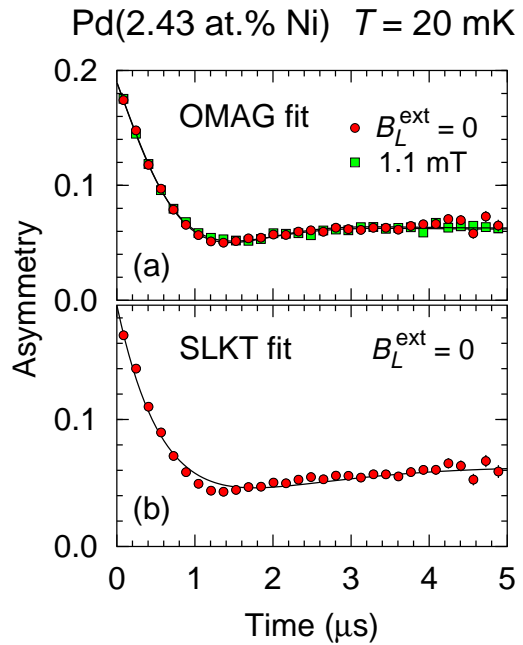}
\caption{\label{fig:kt+omag} (Color online) Asymmetry data from Pd(2.43~at.\,\%~Ni), $T = 20$~mK\@. Circles: external field~$B_L^\mathrm{ext} = 0$ (same data in both panels). (a)~Squares: $B_L^\mathrm{ext} = 1.1$~mT\@, indistinguishable from data for $B_L^\mathrm{ext} = 0$. Curves: fits of OMAG function [Eq.~(\protect\ref{eq:omag})]. (b)~Curve: best fit of static Lorentzian Kubo-Toyabe (SLKT) function [Eq.~(\protect\ref{eq:slkt})] to data for $B_L^\mathrm{ext} = 0$. The fit is noticeably worse than in (a).}
\end{figure}

At 20~mK the data for $B_L^\mathrm{ext} = 0$ and $B_L^\mathrm{ext} = 1.1$~mT, shown in Fig.~\ref{fig:kt+omag}(a), 
are indistinguishable. This is additional evidence for ferromagnetism, since full cancellation by the demagnetizing field is expected for $B_L^\mathrm{ext}$ less than the saturation field~$B^\mathrm{sat}$. The analysis of the zero- and weak longitudinal-field asymmetry data poses a problem, however. As shown in Figs.~\ref{fig:kt+omag}(a) and \ref{fig:kt+omag}(b), the data can be fit by either the OMAG function or the zero-field static Lorentzian Kubo-Toyabe (SLKT) function~\cite{Kubo81,Uemu99} 
\begin{equation}
G_\mathrm{SLKT}(t) = \frac{2}{3}(1-\lambda t) \exp(-\lambda t) + \frac{1}{3} 
\label{eq:slkt}
\end{equation}
appropriate to a Lorentzian distribution of field components~$\langle B^\mathrm{loc}\rangle_i$, $i = x,y,z$, with mean zero and half-width at half-maximum~$\lambda/\gamma_\mu$~\cite{WaWa74}. Both OMAG and SLKT functions fit all data up to $\sim$1.4~K\@. The physical meaning of the two approaches is different, however; the SLKT function is based on a spin-glass-like SRO configuration of the magnetic moments, whereas the nonzero value of $\langle B^\mathrm{loc}\rangle$ indicated by the OMAG function implies an average component over the sample and hence (disordered) LRO\@. The difference between these two limits is essentially whether the correlation length is short or long, and hence is not a sharp distinction.

\begin{figure}[ht]
 \includegraphics[clip=,width=3in]{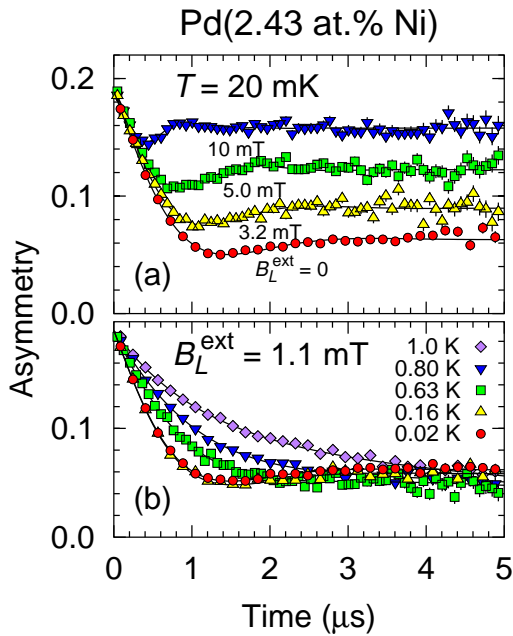}
\caption{\label{fig:20mK+11G} (Color online) Zero-field and weak-longitudinal-field asymmetry data from Pd(2.43~at.\,\%~Ni). (a)~Temperature~$T = 20$~mK, external field~$B_L^\mathrm{ext} = 0$--10~mT\@. (b)~$B_L^\mathrm{ext} = 1.1$~mT, $T = 0.02$--1.0~K\@. Curves: OMAG fits.}
\end{figure}

The OMAG fit in Fig.~\ref{fig:kt+omag}(a) represents the zero-field data noticeably better than the SLKT fit in Fig.~\ref{fig:kt+omag}(b), both visually and in terms of the goodness of fit parameters [reduced $\chi^2 = 1.08$ (OMAG), 1.24 (SLKT)]\@. Nevertheless, a decision as to which of the two approaches is the proper one cannot easily be made solely on the basis of the zero-field asymmetry data. Considering the fact that a 10~mT transverse field enters the sample (Fig.~\ref{fig:tf-2sig}) one might assume that ferromagnetism is not present, but this would not be correct if $B^\mathrm{sat}$ is small. Moreover, data taken in longitudinal fields up to 10~mT, shown in Fig.~\ref{fig:20mK+11G}(a) 
for $T = 20$~mK and discussed further below, give unsatisfactory results when analyzed with the SLKT function, but can be understood if ferromagnetic order is assumed. 

Based on these considerations, together with more conclusive evidence from longitudinal-field data discussed below in Sec.~\ref{sec:2p4lf}, we used the OMAG function [Eq.~(\ref{eq:omag})] to analyze all asymmetry data for $B_L^\mathrm{ext} = 1.1$~mT up to 1~K\@. Data at representative temperatures together with their fits are presented in Fig.~\ref{fig:20mK+11G}(b). 

\begin{figure}[ht]
 \includegraphics[clip=,width=3in]{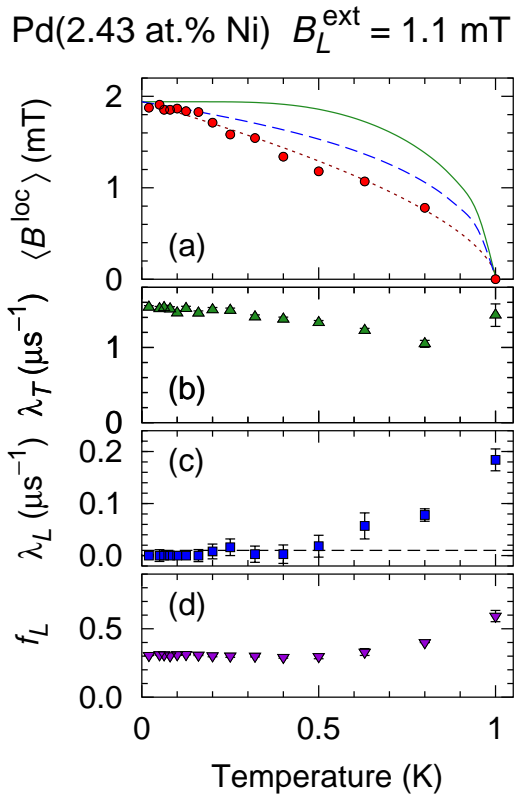}
\caption{\label{fig:pdni24-all} (Color online) Temperature dependencies of parameters from OMAG fits to LF-$\mu$SR data from Pd(2.43~at.\,\%~Ni), $B_L^\mathrm{ext} = 1.1$~mT, $T = 0.02$--1~K\@. (a)~Average local field~$\langle B^\mathrm{loc}\rangle$. Solid curve: mean-field order parameter, $S=1/2$. Dashed curve: mean-field order parameter, $S = \infty$. Dotted curve: power law (see text). (b)~Transverse relaxation rate~$\lambda_T$. (c)~Longitudinal relaxation rate~$\lambda_L$. Dashed line: approximate minimum measurable rate due to muon lifetime. (d)~Fraction~$f_L$ of longitudinal (late-time) polarization.}
\end{figure}

The temperature dependencies of the parameters obtained from these fits are shown in Fig.~\ref{fig:pdni24-all} 
for $T \leq1$~K\@. Figure~\ref{fig:pdni24-all}(a) gives $\langle B^\mathrm{loc}(T)\rangle$. The solid and dashed curves give the mean-field order parameters~\cite{Darb67} for $J = 1/2$ and $J = \infty$ (classical magnetic moments), respectively (for $3d$ transition-metal ions $J = S$). At intermediate temperatures the data lie significantly below both these curves. The dotted curve is a fit of the phenomenological power law~$\langle B^\mathrm{loc}(T)\rangle = \langle B^\mathrm{loc}(0)\rangle[1 - (T/T_C)]^\beta$ to the data, with $\langle B^\mathrm{loc}(0)\rangle = (1.94 \pm 0.02)$~mT, $T_C = (1.00 \pm 0.02)$~K, and $\beta = 0.59 \pm 0.04$. The fit is quite good, but at low temperatures the curve is not as temperature-independent as the data, and the fit value of $\langle B^\mathrm{loc}(T)\rangle$ is somewhat higher than the value~$(1.86 \pm 0.02)$~mT obtained from the OMAG fit at 20~mK [Fig.~\ref{fig:kt+omag}(a)]. The exponent~$\beta$ is not far from the mean-field critical value of 1/2, but should not be considered a critical exponent since data far from $T_C$ were used in the fit. 

Figures~\ref{fig:pdni24-all}(b) and \ref{fig:pdni24-all}(c) give the temperature dependencies of $\lambda_T$ and $\lambda_L$, respectively. The $T=0$ value of $\lambda_T(0) = (1.54 \pm 0.05)~\mu\mathrm{s}^{-1}$ leads to a field distribution width~$\Delta B^\mathrm{loc} = (1.81 \pm 0.06)$~mT\@. Thus $\Delta B^\mathrm{loc} \approx B^\mathrm{loc}$, indicating that the ferromagnetism is strongly disordered. With increasing temperature $\lambda_T$ decreases somewhat but increases again near $T_C$, indicating that a distribution of static or quasistatic fields is still present near the transition. Below $T_C$ $\lambda_L$ decreases rapidly with decreasing temperature, and becomes too small to be measured accurately below ${\sim}0.5~T_C$\@. As expected from the discussion of Sec.~\ref{sec:p5n}, the longitudinal fraction~$f_L$, shown in Fig.~\ref{fig:pdni24-all}(d), is $\approx 1/3$ at low temperatures, and then increases toward 1 as $T \to T_C$ and the static local fields become smaller than $B_L^\mathrm{int}$.

$\mu$SR is a highly local probe, and $\mu$SR data from a spin glass and a strongly disordered weak ferromagnet are quite similar. This is primarily due to the fact that in systems with orientational disorder $\mu$SR is insensitive to whether the disorder is local (spin-glass-like) or macroscopic (domains, grains in a powder). The main difference between a magnet with LRO magnetism and a spin glass is in the distribution of local-field magnitudes, which is more sharply peaked around its average in the former than in the latter. Disorder in an LRO magnet will blur this distinction, and it is quite understandable that in dilute \textit{Pd}Fe alloys both the SLKT and the OMAG functions are able to fit the ZF-$\mu$SR data below the spin freezing temperature. We treat this question in more detail in the next section.

Below $\sim$0.5~K $\lambda_L$ is near or below the minimum measurable value ${\sim}0.01~\mu\mathrm{s}^{-1}$ imposed by the muon lifetime [dashed line in Fig.~\ref{fig:pdni24-all}(c)], indicating that the ferromagnetic spin system is essentially in the static limit. We shall see in Sec.~\ref{p2d4n-HT}, however, that regions of frozen electronic-spin magnetism persist up to $\sim$2~K.

\subsubsection{Longitudinal-Field Data, $T \lesssim T_C$} \label{sec:2p4lf}

The evidence for LRO from the ZF data discussed in Sec.~\ref{sec:zfwftltc} is suggestive but not compelling. Here we consider LF-$\mu$SR data taken at 0.02 and 1~K, which are crucial for the conclusion that LRO is present in the 2.43~at.\,\%~Ni alloy sample. 

\begin{figure*}[ht]
 \includegraphics[clip=,width=\textwidth]{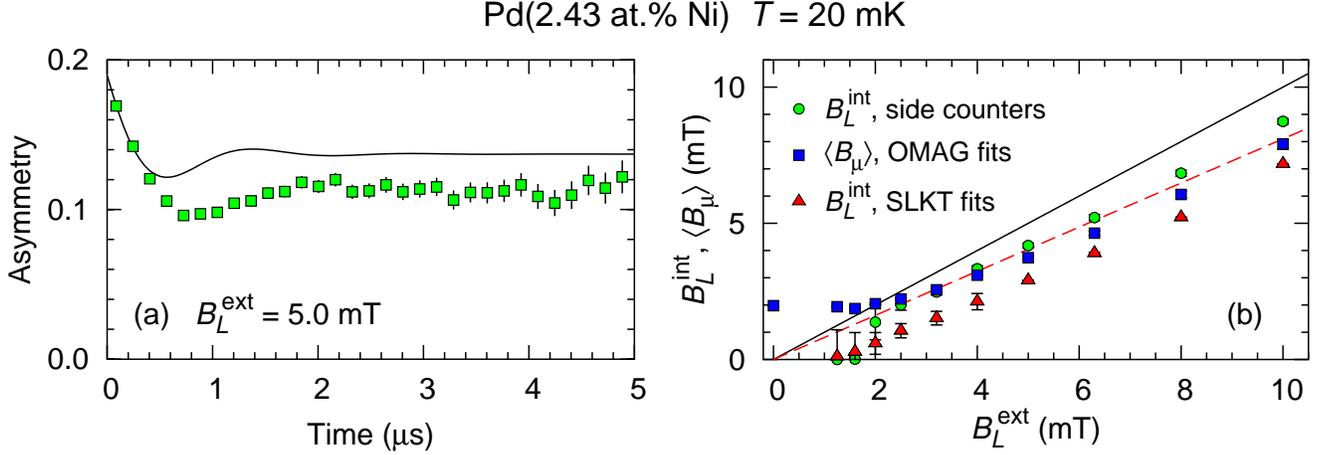}
\caption{\label{fig:2p4-SLKTvsOMAG-field} (Color online) LF-$\mu$SR data from Pd(2.43~at.\,\%~Ni), $T = 20$~mK\@. (a)~ Asymmetry data (points) and SLKT function (curve) for $B_L^\mathrm{ext} = 5.0$~mT\@. (b)~Dependence on external longitudinal field~$B_L^\mathrm{ext}$ of internal field~$B_L^\mathrm{int}$ from side counters (circles) and SLKT fits (triangles), and total static field~$\langle B_\mu\rangle$ from OMAG fits (squares). Solid line: $B_L^\mathrm{int} = B_L^\mathrm{ext}$. Dashed line: $B_L^\mathrm{int} = 0.81B_L^\mathrm{ext}$ (see text).}
\end{figure*}

A proper fit in terms of the longitudinal field ``decoupling'' of the SLKT function~\cite{HUIN79,Uemu99}, discussed briefly in Sec.~\ref{sec:p5n}, could only be achieved by allowing the internal decoupling field~$B_L^\mathrm{int}$ to be a free fit parameter. In the normal Kubo-Toyabe model~\cite{KuTo67,HUIN79} the decoupling field is the external longitudinal field~$B_L^\mathrm{ext}$ and not a fit parameter. It is found that $B_L^\mathrm{int}$ from the SLKT fits is always significantly less than $B_L^\mathrm{ext}$. Figure~\ref{fig:2p4-SLKTvsOMAG-field}(a) 
shows the longitudinal field asymmetry data at 20~mK for $B_L^\mathrm{ext} = 5.0$~mT, for which the fit value of $B_L^\mathrm{int}$ is 2.9~mT (fit not shown). The results for 1~K are quite similar. Any problem with the field controlling electronics was ruled out by test measurements. This unusual behavior is intrinsic to the sample, implying that the SLKT description, and with it a spin-glass-like magnetic ground state, cannot be correct.
 
We therefore analyzed the longitudinal field data using the OMAG function appropriate to LRO, and in addition determined $B_L^\mathrm{int}$ from side-counter precession frequencies as discussed in Sec.~\ref{sec:exp}\@. In Fig.~\ref{fig:2p4-SLKTvsOMAG-field}(b) $B_L^\mathrm{int}$ from side counters and SLKT fits are plotted versus $B_L^\mathrm{ext}$, together with $\langle B_\mu\rangle$ from OMAG fits. It can be seen that the side counters yield $B_L^\mathrm{int} \approx 0$ for $B_L^\mathrm{ext} \lesssim 2$~mT, indicating that Pd(2.43~at.\,\%~Ni) is indeed a ferromagnet, with $B^\mathrm{sat}$ about this value. Above $\sim$2~mT all three internal fields are smaller than $B_L^\mathrm{ext}$, with the SLKT fit values showing the largest discrepancy.

We attribute the reduction of $B_L^\mathrm{int}$ to the demagnetizing field. In general $B^\mathrm{int}$ is given by~\cite{YaDdR11} 
\begin{equation} \label{eq:demag}
B^\mathrm{int} = \left(1 - \frac{4\pi D\chi_V}{1 + 4\pi D\chi_V}\right) B^\mathrm{ext}\,,
\end{equation}
where $D$ is the sample demagnetization factor and $\chi_V$ is the volume susceptibility. 
From measurements on our roughly ellipsoidal Pd(2.43~at.\,\%~Ni) sample, $D \approx 0.75$ in longitudinal field. From Ref.~\cite{CKG81}, in Pd(2.5~at.\,\%~Ni) $\chi_V = 0.024$ at 2.4~K, $4\pi D\chi_V \approx 0.23$, and the internal longitudinal field~$B^\mathrm{int}_L \approx 0.81B_L^\mathrm{ext}$. This should be taken only as a crude estimate, because the susceptibility is presumably different in the present case, but even so a proportionality with this coefficient [dashed line in Fig.~\ref{fig:2p4-SLKTvsOMAG-field}(b)] reproduces the side-counter fields fairly well. The difference between side-counter and SLKT fields (and, to a lesser extent, OMAG fields) cannot be attributed to the demagnetization field, however, and must be thought of as inapplicability of the SLKT fit function. 

Pratt~\cite{Prat07} has treated the case where $\mathbf{B}^\mathrm{loc}$ is constant in magnitude but randomly oriented in a polycrystalline sample. His result for the dependence of the longitudinal signal fraction~$f_L$ on applied longitudinal field~$B_L$ (the ``decoupling curve'') is
\begin{equation}
f_L(b) = \frac{1}{8} + \frac{1}{8b^2} - \frac{(b^2-1)^2}{16b^3}\ln\left| \frac{b+1}{b-1} \right| \,, 
\end{equation}
where $b = B_L/B^\mathrm{loc}$ ($B_L = B_L^\mathrm{int}$ in a ferromagnet). Pratt notes that a general distribution of magnitudes can be accounted for by calculating the average of $f_L(b)$:
\begin{equation} \label{eq:Azavg}
f_L^\mathrm{avg}(B_L) = \int dB^\mathrm{loc}\,P(B^\mathrm{loc})\, f_L(B_L/B^\mathrm{loc}) \,, 
\end{equation}
where $P(B^\mathrm{loc})$ is the distribution function for the local field magnitudes.

When the field components follow Lorentzian distributions with half-width~$\lambda_T$ and zero mean, the corresponding function~$P_\mathrm{Lor}(B^\mathrm{loc})$ for the field magnitudes is
~\cite{Kubo81,UYHS85} 
\begin{equation} \label{eq:PL0}
P_\mathrm{Lor}(B^\mathrm{loc}) = \frac{4}{\pi} \frac{\lambda_T ({B^\mathrm{loc}})^2}{(\lambda_T^2 + ({B^\mathrm{loc}})^2)^2} \,, 
\end{equation}
where fields are given in frequency units ($\gamma_\mu = 1$). With this distribution $f_L^\mathrm{avg}(B_L)$ from Eq.~(\ref{eq:Azavg}) gives the same numerical result as the general formula for the SLKT polarization function~$G_L^\mathrm{Lor}(t)$ in nonzero $B_L$~\cite{Kubo81}.
\begin{eqnarray} \label{eq:Kgolden}
f_L^\mathrm{Lor}(B_L) & = & G_L^\mathrm{Lor}(B_L,t{\to}\infty) \nonumber \\
& = & 1 - \frac{2}{B_L} \int_0^\infty dt'\, \left[ \frac{Q'(t')}{t'} \right] j_0'(B_L t') \,.
\end{eqnarray}
Here $Q(t) = \exp(-\lambda_Tt)$ is the transverse-field relaxation function for a Lorentzian distribution, and $j_0'(x) = \cos x/x - \sin x/x^2$. 

Recognizing that $P_\mathrm{Lor}(B^\mathrm{loc})$ has a nonzero average, we nevertheless reserve the designation~`$\langle B^\mathrm{loc}\rangle$' for cases where the distribution is narrower. An interpolation formula for nonzero $\langle B^\mathrm{loc}\rangle$ can be written simply by generalizing Eq.~(\ref{eq:PL0}) to 
\begin{equation} \label{eq:PL}
P_\mathrm{Lor}^{\,\prime}(B^\mathrm{loc}) = \frac{1}{N} \frac{\lambda_T ({B^\mathrm{loc}})^2}{[\lambda_T^2 + (B^\mathrm{loc}-\langle B^\mathrm{loc}\rangle)^2]^2} \,,
\end{equation}
where 
\begin{widetext}
\begin{equation} \label{eq:Pnorm}
N = \frac{1}{2}\left\{\frac{\langle B^\mathrm{loc}\rangle}{\lambda_T} + \left( 1 + \frac{\langle B^\mathrm{loc}\rangle^2}{\lambda_T^2} \right) \left[ \frac{\pi}{2} + \tan^{-1}(\langle B^\mathrm{loc}\rangle/\lambda_T) \right] \right\}
\end{equation}
\end{widetext}
for normalization. We know of no theoretical justification for Eq.~(\ref{eq:PL}), but it goes to the correct limiting forms for $\langle B^\mathrm{loc}\rangle = 0$ (SLKT) and $\langle B^\mathrm{loc}\rangle \gg \lambda_T$ (``Pratt,'' well-defined $B^\mathrm{loc}$). From our previous discussion we expect this ``broadened OMAG'' form to describe the experimental data.

\begin{figure}[ht]
\includegraphics[clip=,width=3in]{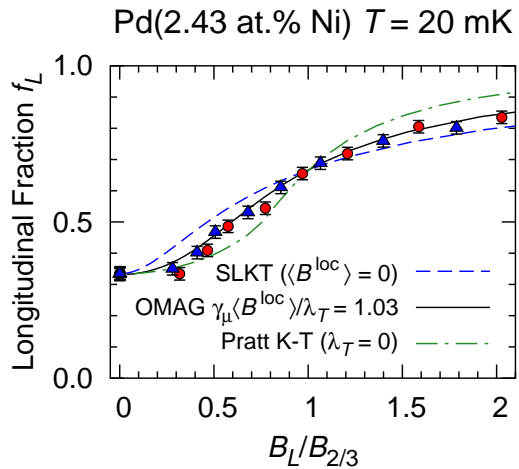}
\caption{\label{fig:scaled} (Color online) Dependence of the longitudinal (late-time) fraction~$f_L$ of the muon polarization on longitudinal field~$B_L$ in Pd(2.43~at.\,\%~Ni), $T = 20$~mK\@. Values of $B_L$ are scaled by $B_{2/3}$, defined by $f_L(B_{2/3}) = 2/3$. Points: $f_L$ from broadened OMAG (circles) and SLKT (triangles) fits. Curves: SLKT (dashed), broadened OMAG (solid), and Pratt (dash-dot) models.}
\end{figure}

Figure~\ref{fig:scaled} 
compares the decoupling curves~$f_L(B_L)$ from the SLKT, broadened OMAG, and Pratt models with experimental SLKT and OMAG fit results to LF-$\mu$SR asymmetry data from Pd(2.43~at.\,\%~Ni) at $T = 20$~mK\@. In order to facilitate the comparison between the model curves and experimental results, $B_L$ for each curve has been scaled by the value~$B_{2/3}$ for which $f_L = 2/3$. For both OMAG and SLKT fits $f_L$ is obtained from the observed late-time asymmetry. In Fig.~\ref{fig:scaled} the values from the OMAG and SLKT fits lie on the same curve, as they should because the late-time behavior does not depend on details of the early-time relaxation. The internal fields from the side counters are used as the experimental values of $B_L$, although the results are not very different if the external fields are used. 

The data clearly lie between the SLKT and Pratt K-T decoupling curves, and are in reasonable agreement with the broadened OMAG prediction for $\gamma_\mu \langle B^\mathrm{loc}\rangle/\lambda_T = 1.03$, the value from the ZF OMAG fit values of $\langle B^\mathrm{loc}(0)\rangle$ and $\lambda_T(0)$. Note that this agreement requires no adjustable parameters. 

It can be seen in Fig.~\ref{fig:scaled} that the slope at $B_L/B_{2/3} = 1$ becomes steeper as $\langle B^\mathrm{loc} \rangle/\lambda_T$ increases, i.e., the relative width of the distribution of $B^\mathrm{loc}$ becomes smaller. Pratt noted~\cite{Prat07} that this shape dependence could be a useful complement to the early-time depolarization functions in determining the form of the microscopic field distribution, and we have analyzed our data in this spirit. 

\subsubsection{Zero- and Longitudinal-Field Data, $T \gtrsim T_C$} \label{p2d4n-HT}

\begin{figure}[ht]
 \includegraphics[clip=,width=3in]{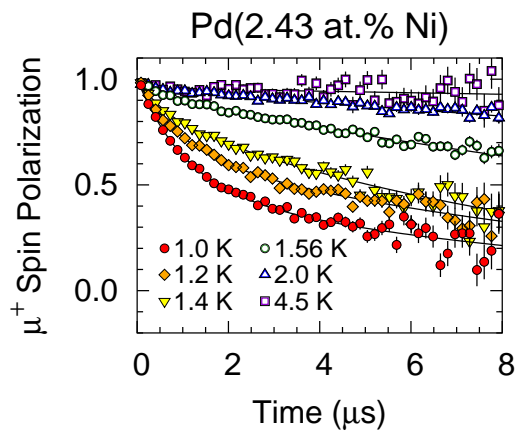}
\caption{\label{fig:zf-ht-spec} (Color online) ZF- and weak LF-$\mu$SR relaxation functions [asymmetry data normalized to $A(0)$] from the 2.43~at.\,\%~Ni alloy at and above $T_C$. 
Filled symbols ($T \le 1.4$~K): LTF spectrometer, $B_L^\mathrm{ext} = 1.1$~mT\@. Open symbols ($T > 1.4$~K): GPS spectrometer, $B_L^\mathrm{ext} = 0$. Solid curves: fits of Eq.~(\protect\ref{eq:twoex})  to the data.} 
\end{figure}

\begin{figure}[ht]
 \includegraphics[clip=,width=3in]{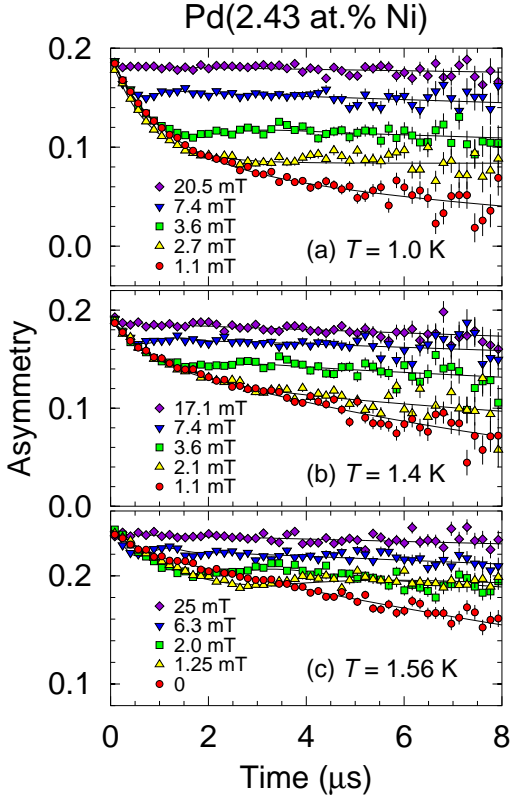}
\caption{\label{fig:HTfield-pol} (Color online) LF-$\mu$SR asymmetry data in Pd(2.43~at.\,\%~Ni). (a)~$T = 1.0$~K\@. (b)~$T = 1.4$~K\@. (c)~$T = 1.56$~K\@. Curves: fits of Eq.~(\protect\ref{eq:twoex})  to the data.}
\end{figure}

ZF-$\mu$SR in the paramagnetic state of a homogeneous ferromagnet above $T_C$ is expected to be dominated by exponential dynamic muon spin relaxation:
\begin{equation}
 G_p(t) = \exp(-\lambda_p t) \,,
\label{pmag}
\end{equation}
 where the relaxation rate $\lambda_p$ is inversely proportional to the fluctuation rate~$\nu_f$ of the paramagnetic spins in the ``motionally narrowed'' limit~$(\lambda_p/\nu_f)^{1/2} \ll 1$~\cite{YaDdR11}. This is not what we observe in Pd(2.43~at.\,\%~Ni), however. Asymmetry data from the 2.43~at.\,\%~Ni alloy for temperatures in the neighborhood of $T_C$ are shown in Fig.~\ref{fig:zf-ht-spec}. 
The decay is not exponential at and above $T_C$, although it becomes more nearly so at 1.4~K and above. A sum of rapidly- and slowly-relaxing exponentials:
\begin{equation}
G(t) = (1-f_L)\exp(-\lambda_Lt) + f_L\exp(-\lambda_Tt)
\label{eq:twoex}
\end{equation}
provides good fits to the data over the entire temperature range $T \ge T_C$. Figure~\ref{fig:HTfield-pol} 
gives LF-$\mu$SR asymmetry data at representative fields for $T = 1.0$~K, 1.4~K, and 1.56~K\@. The observed decoupling is evidence that the initial relaxation is static or quasistatic in origin, i.e., that static electronic magnetism persists above $T_C$\@. 

\begin{figure}[ht]
 \includegraphics[clip=,width=3in]{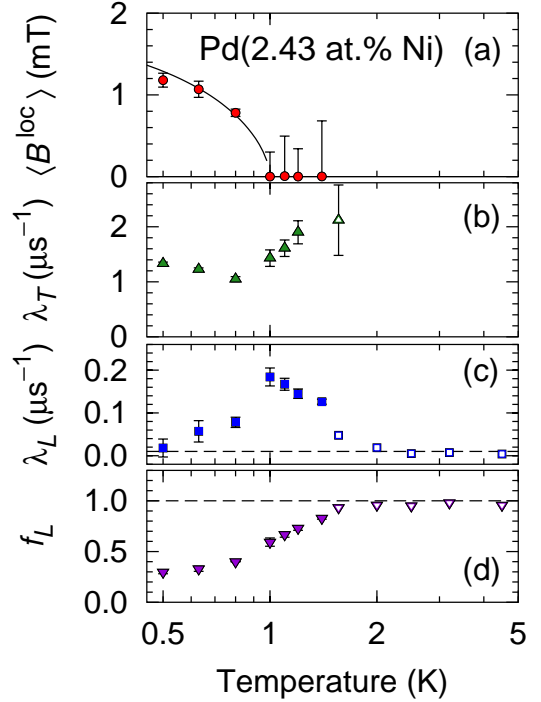}
\caption{\label{fig:pdni24-HTall} (Color online) Temperature dependences in the neighborhood of $T_C$ of parameters from OMAG fits to weak LF- and ZF-$\mu$SR data from Pd(2.43~at.\,\%~Ni). Filled symbols ($T \le 1.4$~K): LTF spectrometer, $B_L^\mathrm{ext} = 1.1$~mT\@. Open symbols ($T > 1.4$~K): GPS spectrometer, $B_L^\mathrm{ext} = 0$. (a)~Average local field~$\langle B^\mathrm{loc}\rangle$. (b)~Transverse relaxation rate~$\lambda_T$. (c)~Longitudinal relaxation rate~$\lambda_L$. Dashed line: minimum measurable rate due to muon lifetime. (d)~Fraction~$f_L$ of longitudinal (late-time) polarization.}
\end{figure}

Figure~\ref{fig:pdni24-HTall} 
gives the temperature dependence of the parameters obtained from these fits for $T$ in the range 1--5~K\@. Results for $0.5~\mathrm{K} \leq T \leq 1$~K from Fig.~\ref{fig:pdni24-all} are also shown. Surprisingly, $\lambda_T$ increases sharply above 1~K; in a paramagnetic state with no static magnetism both the mean~$\langle B^\mathrm{loc}(T)\rangle$ and the static contribution to $\lambda_T$ would vanish. Clearly, static magnetism persists to $\sim$1.4~K, above which the early-time fraction of the signal is too small to determine $\lambda_T$. 

Longitudinal relaxation becomes considerably stronger in the neighborhood of $T_C$, where $\lambda_L$ increases with decreasing temperature and goes through a cusp at ${\sim}T_C$. This behavior is expected from slowing down of critical spin fluctuations as $T_C$ is approached, but critical dynamics are not normally accompanied by static magnetism above $T_C$. The weak relaxation above ${\sim}2T_C$ is due to rapid Ni spin fluctuations (strong motional narrowing), whereas below ${\sim}0.5T_C$ the weak dynamic relaxation [Fig.~\ref{fig:pdni24-all}(c)] reflects static or nearly static Ni spins.\footnote{By ``Ni spins'' we mean the atomic spins together with any associated Pd spin polarization.}

Figure~\ref{fig:pdni24-HTall}(d) shows that the longitudinal fraction~$f_L$ increases with increasing temperature through $T_C$. We consider two scenarios for this behavior. In the ``inhomogeneous'' picture the disorder is meso- or macroscopic in scale, with separate SRO and paramagnetic regions. Muon spins in the SRO fraction exhibit static or quasistatic relaxation, whereas muons in the paramagnetic fraction are dynamically relaxed, and $f_L(T)$ represents the increase in paramagnetic volume fraction with increasing temperature. A 100\% paramagnetic fraction is reached between 1~K and 2~K, above which a single exponential fit (with a very small relaxation rate) suffices. Alternatively, in the ``homogeneous'' scenario, the scale of the disorder is microscopic, and the muon sites are statistically equivalent. Then $f_L(T)$ is due to decoupling by $B_L^\mathrm{ext} = 1.1$~mT as $\langle B^\mathrm{loc}(T)\rangle$ falls below this value. For temperatures in the range of the GPS (open symbols in Fig.~\ref{fig:pdni24-HTall}), the sample is entirely paramagnetic, and $f_L$ is essentially unity. 

It is not easy to distinguish between these pictures. The observed decoupling up to 1.56~K (Fig.~\ref{fig:HTfield-pol}) is evidence that the initial relaxation is static, and the data are consistent with this static magnetism occupying the entire sample volume. A necessary condition for the homogeneous scenario is that the increase of $f_L(T)$ begins at the temperature where $B^\mathrm{loc}(T)$ falls below $B_L^\mathrm{ext}$. From Figs.~\ref{fig:pdni24-HTall}(a) and \ref{fig:pdni24-HTall}(d) it can be seen that this is roughly the case. For example, at 1.4~K $\lambda_T/\gamma_\mu B^\mathrm{loc} \sim 2$ and $f_L \approx 0.85$, consistent with the decoupling curves of Fig.~\ref{fig:scaled}. 

This is, however, not sufficient to establish the homogeneous scenario, and support for large-scale inhomogeneity comes from the fact that $\lambda_T$ shows no sign of decreasing above $T_C$. Furthermore, as discussed below in Sec.~\ref{sec:p2d5n-HT}, in the 2.5~at.\,\%~Ni alloy the behavior above $T_C \approx 2$~K is essentially the same as described above, except scaled to higher temperatures. Since these data were taken in the GPS with $B_L^\mathrm{ext} = 0$, the homogeneous scenario for $f_L(T)$, which requires an external longitudinal field, is not applicable. This in turn suggests the inhomogeneous picture for the Pd(2.43~at.\,\%~Ni) sample. We conclude that segregation of SRO and paramagnetic regions most likely sets in at ${\sim}T_C$ for both samples, with a rapidly-decreasing SRO volume fraction with increasing temperature. 

\subsection{\label{sec:p2p5n} Pd(2.5~at.\,\%~Ni)} 

Data were taken from this sample over the temperature range~0.02--30~K using both the LTF and the GPS\@. Since the sample was not particularly large and the LTF spectrometer is not equipped with a veto system, a background signal was present in the LTF\@. In order to determine this background signal accurately, we mounted the 2.5~at.\,\%~Ni alloy button on pure Pd foils. In transverse field measurements the background signal was seen as an almost undamped oscillation.

\subsubsection{Transverse-Field Data}

\begin{figure}[ht]
 \includegraphics[clip=,width=3in]{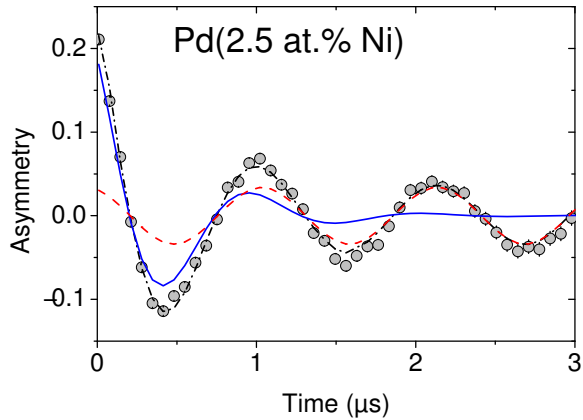}
 \caption{\label{tf-2sig} (Color online) TF-$\mu$SR asymmetry in the 2.5~at.\,\%~Ni alloy at 20~mK and $B_T^\mathrm{ext} = 7$~mT, fit to the sum of sample and background signals. Solid curve: sample signal; dashed curve: background signal; dash-dot curve: sum signal. For details see text.}
\end{figure}

Asymmetry data taken at 20~mK in a transverse field~$B_T^\mathrm{ext} = 7$~mT are shown in Fig.~\ref{tf-2sig}, 
together with a fit to the sum of damped (sample) and undamped (background) oscillatory signals. The background signal contributes about 12\% of the total asymmetry, and has been subtracted for further analysis of the zero- and longitudinal-field data. The existence of an oscillatory signal from the sample shows that $B_T^\mathrm{ext}$ enters the bulk of the material. The field at the muon sites in the sample is $\sim$2.5\% lower than in the pure Pd foils surrounding the sample. This is probably due to strong demagnetization, since $B_T^\mathrm{ext}$ may not be much larger than $B^\mathrm{sat}$. 

\subsubsection{Zero- and Longitudinal-Field Data, $T < T_C$}

The analysis of ZF- and LF-$\mu$SR data from the 2.5~at.\,\%~Ni alloy was carried out as for the 2.43~at.\,\%~Ni alloy. One encounters the same situation as before: the zero field data can be fit satisfactorily by either the SLKT or the OMAG function but, as for the 2.43~at.\,\%~Ni alloy, the LF-$\mu$SR data cannot be reproduced properly within the SLKT model.

\begin{figure}[ht]
 \centering
 \includegraphics[clip=,width=3in]{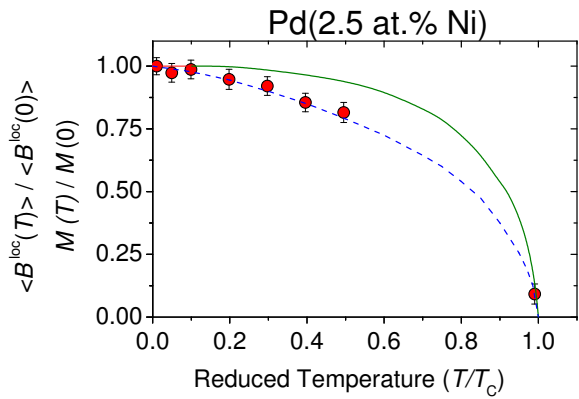}
\caption{\label{bmu} (Color online) Data: reduced local field vs reduced temperature for Pd(2.5~at.\,\%~Ni) (filled points). Curves: mean-field magnetization for $S=1/2$ (solid curve) and $S= \infty$ (dashed curve).}
\end{figure}

We therefore used the OMAG function to analyze all spectra up to 2~K with complete success. A zero-temperature value of $\langle B^\mathrm{loc}(0)\rangle = (3.8 \pm 0.1)$~mT was found. At 2~K the OMAG fit returned the very small value~$\langle B^\mathrm{loc}\rangle = (0.3 \pm 0.15)$~mT\@. Extrapolating smoothly to $\langle B^\mathrm{loc}\rangle = 0$ results in $T_C = (2.03 \pm 0.03)$~K\@. In Fig.~\ref{bmu} 
the reduced local field~$\langle B^\mathrm{loc}(T)\rangle/\langle B^\mathrm{loc}(0)\rangle$ is plotted vs. reduced temperature $T/T_C$, together with the mean field magnetization curves for ferromagnets with $S=1/2$ and $S=\infty$. 

For the transverse relaxation rate the fits yield $\lambda_T = (3 \pm 0.5)~\mu\mathrm{s}^{-1}$ independent of temperature, which gives a field distribution 
width~$\Delta B_\mu \approx 3.5$~mT. This value is comparable to the saturation value of the mean local field, meaning that here as well the ferromagnetic spin structure is strongly disordered. The resulting strong damping again prevents the development of a full oscillatory pattern. For all temperatures below $T_C$ the fit values of the longitudinal relaxation rate are well below the minimum measurable value of ${\sim}0.01~\mu\mathrm{s}^{-1}$. This means that immediately below $T_C$ the ferromagnetic spin system is in the static limit, a quite abnormal behavior. The difference between this behavior and the observed maximum in $\lambda_L(T)$ near $T_C$ in the 2.43~at.\,\%~Ni alloy [Fig.~\ref{fig:pdni24-HTall}(c)] may be due to considerably faster spin fluctuation rates in the 2.5~at.\,\%~Ni sample, leading to motional narrowing and suppression of $\lambda_L$.

\subsubsection{Zero-field data above $T_C$} \label{sec:p2d5n-HT}

\begin{figure}[ht]
 \includegraphics[clip=,width=3in]{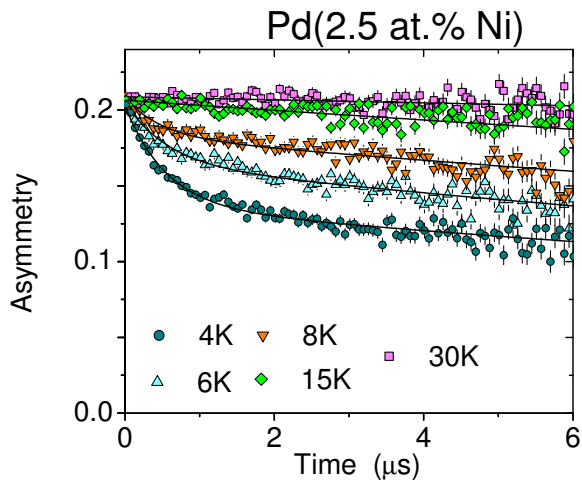}
\caption{\label{fig:2p5-zf-lt-spec} (Color online) ZF-$\mu$SR asymmetry data from the 2.5~at.\,\%~Ni alloy above 2~K\@. The spectra at 4, 6, and 8~K were fitted with the sum of rapidly- and slowly-relaxing exponential components; the data at 15 and 30~K with a slowly relaxing exponential only. The data were obtained using the GPS spectrometer. }
\end{figure}

\begin{figure}[ht]
 \includegraphics[clip=,width=3in]{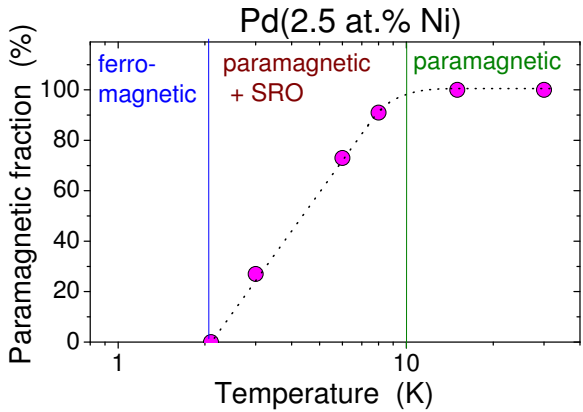}
\caption{\label{fig:phdia} (Color online) Schematic phase diagram of the magnetic states of Pd (2.5~at.\,\%~Ni).}
\end{figure}

As noted in Sec.~\ref{p2d4n-HT}, in a homogeneous ferromagnet one expects to observe paramagnetic behavior above $T_C$, i.e., only dynamic relaxation, which is described by an exponential relaxation function. Asymmetry data from the 2.5~at.\,\%~Ni alloy above 2~K are shown in Fig.~\ref{fig:2p5-zf-lt-spec}. 
It is apparent that for the asymmetry data at 4, 6, and~8~K a single exponential fit is not appropriate. As for the 2.43~at.\,\%~Ni alloy (Sec.~\ref{p2d4n-HT}), the data are well fit by the sum of two exponential components, one rapid and one slow. It is again found that the fraction of the slow component increases with increasing temperature. A 100\% paramagnetic fraction is reached between 8 and 15~K, where a single exponential fit suffices. By interpreting the relative intensity of the rapidly-relaxing signal as the relative volume fraction of a SRO state, and accordingly that of the slowly-relaxing signal as the relative volume fraction of the paramagnetic state, one obtains the $\mu$SR-based schematic magnetic phase diagram shown in Fig.~\ref{fig:phdia}.

Above $T_C$ the relaxation rate of the paramagnetic fraction is very low, i.e., Ni spin fluctuations are rapid. No increase in rate is observed when approaching $T_C$ from above. This, together with the evidence noted above that the ferromagnetic spin system enters the static limit immediately below $T_C$, suggests that, unlike the situation in Pd(2.43~at.\,\%~Ni), the transition is mean-field-like, i.e., without critical fluctuations. Equivalently, the dynamic critical region is either too narrow to be visible or is obscured by a distribution of transition temperatures.

The ZF-$\mu$SR experiments on the 2.5~at.\,\%~Ni alloy differ from those on the 2.43~at.\,\% Ni sample in that all spectra for $T \geq 2$~K were taken with the GPS spectrometer in true ZF, so that there is no possibility of the homogeneous ``decoupling'' scenario that requires an external field. Thus the data indicate that the rapidly-relaxing component is due to a quasistatic SRO fraction, with the slowly relaxing component due to a free paramagnetic fraction, i.e., the sample is inhomogeneous on a length scale (longer than a few lattice parameters) such that a given muon relaxes rapidly or slowly depending on which fraction it occupies. This in turn suggests that the same scenario is applicable to the 2.43~at.\,\%~Ni alloy.

\section{\label{sum} Discussion and Summary} 

\paragraph{Muon diffusion?} Muon diffusion in pure Pd combined with trapping-detrapping effects at the Ni impurities~\cite{HHWH92,Karl95} are possible complications of the $\mu$SR studies. However, the onset below $T_C$ of a static field~$\langle B^\mathrm{loc}\rangle$ in the 2.4 and 2.5~at.\,\%~Ni alloys is qualitatively similar to the behavior in the well-established ferromagnet Pd(5~at.\,\%~Ni), so that muon diffusion does not appear to be appreciable in any of these alloys. Quenching of muon diffusion by disorder is a common phenomenon~\cite{Sche85,Karl95,YaDdR11}.

\paragraph{Comparison of alloys close to $x_\mathrm{cr}$.} $\mu$SR properties from experiments in the 2.43~at.\,\%~Ni and 2.5~at.\,\%~Ni alloys are summarized in Table~\ref{tbl:summary}.
\begin{table}[ht]
\caption{\label{tbl:summary} Curie temperature~$T_C$, crossover temperature $T_p$ to paramagnetic region, $T{=}0$ mean local field $\langle B^\mathrm{loc}(0)\rangle$, $T{=}0$ transverse relaxation rate~$\lambda_T(0)$, ratio $\gamma_\mu\langle B^\mathrm{loc}(0)\rangle/\lambda_T(0)$, and experimental and calculated values of the product $p_\mathrm{eff}c$ in Pd$_{1-x}$Ni$_x$, $x = 0.0243$ and 0.025.}
\begin{ruledtabular}
\begin{tabular}{ldd}
$x$ & 0.0243 & 0.025 \\
\colrule
$T_C~(\mathrm{K})$\footnotemark[1] & 1.00 \pm 0.02 & 2.03 \pm 0.03 \\
$T_p~(\mathrm{K})$\footnotemark[2] & {\sim}2 & {\sim}12 \\
$\langle B^\mathrm{loc}(0)\rangle~(\mathrm{mT})$\footnotemark[3] & 1.86 \pm 0.02 & 3.8 \pm 0.1 \\
$\lambda_T(0)~(\mu\mathrm{s}^{-1})$\footnotemark[3] & 1.54 \pm 0.05 & 3.0 \pm 0.5 \\
$\gamma_\mu\langle B^\mathrm{loc}(0)\rangle/\lambda_T(0)$ & 1.03 \pm 0.03 & 1.1 \pm 0.2 \\
$(p_\mathrm{eff}c)_\mathrm{exp}$\footnotemark[4] & 0.010 & 0.019 \\
$(p_\mathrm{eff}c)_\mathrm{calc}$\footnotemark[5] & 0.022 & 0.024 \\
\end{tabular}
\footnotetext[1]{From $\langle B^\mathrm{loc}(T_C)\rangle = 0$ (Sec.~\protect\ref{sec:zfwftltc}).}
\footnotetext[2]{From $f_L = 1$, cf.\ Figs.~\protect\ref{fig:pdni24-HTall}(d) and \protect\ref{fig:phdia}.}
\footnotetext[3]{From OMAG fits to ZF data, $T = 20$~mK.}
\footnotetext[4]{From Eqs.~(\protect\ref{eq:peffcexp}) and (\protect\ref{eq:lambdaprime}).}
\footnotetext[5]{From Eq.(\protect\ref{eq:peffccalc}), using values from Ref.~\protect~\cite{CKG81}.}
\end{ruledtabular}
\end{table}
The values of $T_C$ are in good agreement with previous reports~\cite{MTC74}. Since these alloys have Ni concentrations very close to the critical value and their $\mu$SR properties are similar, they can be discussed together. The lower values of $T_C$, $\langle B^\mathrm{loc}(0)\rangle$, and $\lambda_T(0)$ in the 2.43~at.\,\%~Ni alloy compared to Pd(2.5~at.\,\%~Ni) are in agreement with the lower Ni concentration. It is remarkable, however, that these quantities differ by a factor of 2, while the change in Ni content is quite small. This high sensitivity of magnetic parameters to the Ni concentration indicates that we must indeed be close to $x_\mathrm{cr}$ but not below it. From the data in Table~\ref{tbl:summary} a linear extrapolation of $T_C(x)$ to $T_C(x_\mathrm{cr}) = 0$ yields $x_\mathrm{cr} = 0.0236 \pm 0.0027$, in reasonable agreement with values derived from resistivity and magnetic measurements~\cite{CrSc65,TaCo71,MTC74,CKG81}. 

\paragraph{Ni clusters for $T \ll T_C$?} Both dilute samples are weak ferromagnets, since they can be easily magnetized in low external fields even in polycrystalline form. The muon local field is strongly disordered, reflecting corresponding disorder in the spin system. The saturation values of the local fields are extremely low. 

We next consider implications of our data for the magnitudes and distribution of static magnetic moments in these alloys. If we consider a model with randomly-oriented static moments on all Ni sites, the spin-glass calculation of $\mathbf{B}^\mathrm{loc}$ by Uemura \textit{et al.}~\cite{UYHS85} is appropriate. The Lorentzian distribution of field components~\cite{WaWa74} leads to the result
\begin{equation}
\lambda_T = (\pi/2)^{1/2} c\, \Delta_\mathrm{max} \,,
\end{equation}
where $c$ is the concentration of magnetic impurities and $\Delta_\mathrm{max}/\gamma_\mu$ is the width of the (Gaussian) distribution of $\mathbf{B}^\mathrm{loc}$ when all lattice sites are occupied by magnetic ions; $\Delta_\mathrm{max}$ is due to the dipole-dipole interaction and scales as $p_\mathrm{eff}/a^3$, where $p_\mathrm{eff}$ is the effective moment in Bohr magnetons and $a$ is the lattice parameter~\cite{UYHS85}. For \textit{Cu}Mn alloys with an effective moment of $5\mu_B$, $\Delta_\mathrm{max} = 1400~\mu\mathrm{s}^{-1}$ assuming an octahedral muon site in the fcc Cu lattice~\cite{UYHS85}. Pd metal is also fcc; the muon site is unknown, but we assume it to be the octahedral site as in Cu. Scaling the value of $\Delta_\mathrm{max}$, we find
\begin{equation} \label{eq:peffcexp}
\lambda_T\ (\mu\mathrm{s}^{-1}) = 231 p_\mathrm{eff}c \quad \mathrm{(Pd)} \,.
\end{equation}
In the ferromagnetic case $\gamma_\mu\langle B^\mathrm{loc}\rangle$ must be considered in addition to $\lambda_T$, as discussed in Sec.~\ref{sec:2p4lf}. From Table~\ref{tbl:summary} these two quantities are comparable in the 2.43~at.\,\% Ni and 2.5~at.\,\%~Ni alloys. Thus we use 
\begin{equation} \label{eq:lambdaprime}
\lambda'_T = (\lambda_T^2 + \gamma_\mu^2\langle B^\mathrm{loc}\rangle^2)^{1/2}
\end{equation}
 as a crude estimate to obtain $(p_\mathrm{eff}c)_\mathrm{exp}$ from Eq.~(\ref{eq:peffcexp}). Values are given in Table~\ref{tbl:summary}. 

Assigning a moment to each Ni atom ($c \approx x$) leads to $p_\mathrm{eff} = 0.5\mbox{--}1\mu_B$. We note, however, that in good solid solutions such as PdNi statistical clustering is inevitable. There is considerable evidence from a number of studies~\cite{BeTo76,Chou76,SaKo78,CKG81,KiGa91} that isolated Ni atoms are nonmagnetic and only Ni clusters become ferromagnetic, and that the effective moment per cluster is much larger than $1~\mu_B$. From their detailed investigation of magnetic properties of dilute PdNi alloys, Kouvel and co-workers~\cite{SaKo78,CKG81} concluded that (1)~for Ni concentration~$x \lesssim 1.8$~at.\,\% only statistical Ni clusters containing 3 or more nearest-neighbor nickel atoms, with concentration~$c_{3+}$, bear moments with $p_\mathrm{eff}^{3+} \approx 17~\mu_B$~\cite{SaKo78}, and (2)~for Ni concentrations in the range~$\sim$1.8--3~at.\,\% Ni some pairs of Ni atoms, with concentration~$c_x$, are also magnetic with moments $p_\mathrm{eff}^x \approx 12.2~\mu_B$~\cite{CKG81}. At $\sim$2.5~at.\,\%~Ni approximately half the pairs are magnetic: $c_x \approx 0.5c_2$, where $c_2$ is the concentration of pairs sharing one and only one nearest-neighbor bond. It has been concluded from neutron studies of ferromagnetic PdNi alloys~\cite{ARS70,BCLR82} that the concentration of polarization clouds is less than that of the Ni atoms. Hence one deals with statistical Ni clusters in PdNi.

Using the values reported in Ref.~\cite{CKG81}, which assumes only statistical Ni clustering, calculated values of the ``effective'' value
\begin{equation} \label{eq:peffccalc}
(p_\mathrm{eff}c)_\mathrm{calc} = p_\mathrm{eff}^{3+}c_{3+} + p_\mathrm{eff}^xc_x
\end{equation}
are given in Table~\ref{tbl:summary}. Two features stand out: the experimental and calculated values are comparable, but the concentration dependence of $(p_\mathrm{eff}c)_\mathrm{calc}$ is considerably smaller than that of $(p_\mathrm{eff}c)_\mathrm{exp}$. We conclude that the $\mu$SR data are in general agreement with the cluster picture, but that the details of cluster magnetism are not captured quantitatively. 

For both dilute alloys the spontaneous muon local field~$\langle B^\mathrm{loc}(T)\rangle$ below $T_C$ falls below the mean-field prediction for low spin values. For the 2.5~at.\,\%~Ni alloy $\langle B^\mathrm{loc}(T)\rangle$ resembles the classical ``Langevin'' mean-field magnetization curve (Fig.~\ref{bmu}). This could be additional evidence for clustering, since large total cluster moments would be expected to lead to classical behavior. For Pd(2.43~at.\,\%~Ni) $\langle B^\mathrm{loc}(T)\rangle$ falls below the classical limit [Fig.~\ref{fig:pdni24-all}(a)]. This may be an indication that a mean-field picture does not hold here or, alternatively, that the additional loss of static magnetism with increasing temperature is due to ``shedding'' of spins by the clusters as $T_C$ is approached. Such behavior might be expected if Ni-Ni exchange couplings are broadly distributed, with a significant portion of the weaker couplings smaller than $k_ BT_C$.

\paragraph{Cluster formation; percolation for $T \gtrsim T_C$.}  In the transition region above $T_C$, the $\mu$SR data suggest separate regions of SRO (fast relaxing signal) and paramagnetism (slowly relaxing signal). This is again an indication of clustering of Ni in the Pd matrix. Small-angle neutron scattering on the 2.5~at.\,\%~Ni alloy also indicates that clustering is present~\cite{Pfle12pc, *PBFK10}. A similar picture has been invoked for the magnetic clusters formed by the giant moments in Fe doped Pd~\cite{KSP95}. We also note that the spread of coexistence of SRO and paramagnetic $\mu$SR signals is narrower in the 2.43~at.\,\%~Ni alloy ($\sim 2T_C$) than in the 2.5~at.\,\%~Ni alloy ($\sim 4T_C$). This indicates that besides a lower number of Ni clusters, the clusters have smaller volumes at lower Ni concentration. Such a tendency is partially captured by randomly-formed clusters, but not with the rather large difference found experimentally.

Below $T_C$, where the magnetic moments are fully correlated, only one $\mu$SR signal is seen with full intensity, implying that matrix and clusters must be treated here as a single entity. We noted in Sec.~\ref{sec:zfwftltc} that in the presence of considerable short-range disorder the main difference between LRO and a spin glass is the existence of an average spontaneous magnetization in the former. \textit{Pd}Fe and \textit{Pd}Mn alloys form spin-glass ground states, the latter exhibiting ferromagnetism at low concentrations~\cite{HMW82,PBKM84}. The different behavior of the PdNi system might be due to details of the percolation process~\cite{Odod83,StWi00}, perhaps associated with the lack of moment on isolated Ni atoms.

We have noted that statistical clustering is always present in solid solutions such as PdNi. It is often assumed that for the existence of a QCP at $x_\mathrm{cr}$ a homogeneous alloy is needed. However, magnetic, thermal and resistivity measurements~\cite{NBKM99} carried out on the identical sample of Pd(2.5~at.\,\%~Ni) as used in the present $\mu$SR work have shown that this sample clearly exhibits non-Fermi-liquid behavior that is commonly taken as an indicator for quantum critical behavior. This raises the question of whether homogeneity in Ni distribution is a stringent condition for a QCP or, alternatively, if NFL behavior reflecting a QCP is also a characteristic of the cluster state. In this regard, it should be noted that a percolation approach also generates non-Fermi-liquid values of transport exponents~\cite{StWi00}. 

Our results do not give a definitive answer to the question of whether a quantum critical point is present. $\mu$SR experiments concerned with a QCP are scarce; a recent example is a study of CeRhSi$_3$~\cite{EGMG12}. This heavy fermion antiferromagnet (AFM) becomes superconducting at pressures $P > 12$~kbar, with AFM vanishing at $P_\mathrm{cr} = 23.6$~kbar. This loss of LRO is considered a magnetic QCP\@. The $\mu$SR data show that $T_C(P)$ and $\langle B^\mathrm{loc}(P)\rangle$ both vanish at $P_{\rm cr}$. Although we did not reach $x_\mathrm{cr}$ exactly, there is evidence for such disappearance in our data, since $(p_\mathrm{eff}c)_\mathrm{exp}$ decreases more rapidly than expected as $x \to x_\mathrm{cr}$ (Table~\ref{tbl:summary}). 

No information on spin dynamics is given in Ref.~\cite{EGMG12}. The present study suggests that for $x = 0.0243$, slightly above $x_\mathrm{cr}$, Ni spin fluctuations exhibit critical slowing down as $T_C$ is approached from above, i.e., more or less the expected behavior for a normal transition. For $x = 0.025$ the Ni spin fluctuations are apparently too rapid to relax muon spins in the available time window. The absence of critical slowing down together with the mean-field behavior of $\langle B^\mathrm{loc}\rangle$ suggest a mean-field-like transition in this alloy. All spins exhibit a static component immediately below $T_C$, as in an ordinary transition. 

Further work, both theoretical and experimental, is needed to understand the curious properties of the PdNi system. In particular, it would be desirable to study alloys with less than 2.3~at.\,\%~Ni, in order to approach $x_\mathrm{cr}$ from below. Future work should also explore the Griffiths-phase~\cite{Grif69} scenario for the ``paramagnetic + SRO'' region of Fig.~\ref{fig:phdia}, as discussed in, e.g., Ref.~\cite{Vojt10}. 

{\it Note added in proof}: We have become aware of an alternative treatment of quasistatic muon relaxation in a partially-ordered internal field by Larkin {\it et al.}, Physica B {\bf 289-290}, 153 (2000), which we find reproduces the results of our OMAG analysis with only minor quantitative differences.

\begin{acknowledgments}
We thank H. L\"utkens and R. Scheuermann (Swiss Muon Source) and R.~H. Heffner (Los Alamos) for their help in carrying out the experiments. This work was partially supported by the Deutsche Forschungsgemeinschaft (DFG) via TRR80 (Augsburg, Munich, Stuttgart) and FOR 960, and by the U.S. NSF, Grant Nos.~DMR-9731361 and DMR-0102293 (UC Riverside) and DMR-9820631 and DMR-1105380 (CSU Los Angeles). Work at Brookhaven National Laboratory was carried out under the auspices of the U.S. Department of Energy, Office of Basic Energy Sciences under Contract No.~DE-AC02-98CH1886. Research at U.C. San Diego was supported by the U.S. Department of Energy under Grant No.~DE-FG02-04-ER46105.

This paper is dedicated to Gerard J. Nieuwenhuys, who passed away in 2010. He was an outstanding physicist, a valued collaborator, and our dear friend.
\end{acknowledgments}


\begin{thebibliography}{10}%
\makeatletter
\providecommand \@ifxundefined [1]{%
 \ifx #1\undefined \expandafter \@firstoftwo
 \else \expandafter \@secondoftwo
\fi
}%
\providecommand \@ifnum [1]{%
 \ifnum #1\expandafter \@firstoftwo
 \else \expandafter \@secondoftwo
\fi
}%
\providecommand \enquote [1]{``#1''}%
\providecommand \bibnamefont  [1]{#1}%
\providecommand \bibfnamefont [1]{#1}%
\providecommand \citenamefont [1]{#1}%
\providecommand\href[0]{\@sanitize\@href}%
\providecommand\@href[1]{\endgroup\@@startlink{#1}\endgroup\@@href}%
\providecommand\@@href[1]{#1\@@endlink}%
\providecommand \@sanitize [0]{\begingroup\catcode`\&12\catcode`\#12\relax}%
\@ifxundefined \pdfoutput {\@firstoftwo}{%
 \@ifnum{\z@=\pdfoutput}{\@firstoftwo}{\@secondoftwo}%
}{%
 \providecommand\@@startlink[1]{\leavevmode}%
 \providecommand\@@endlink[0]{}%
}{%
 \providecommand\@@startlink[1]{%
  \leavevmode
  \pdfstartlink
   attr{/Border[0 0 1 ]/H/I/C[0 1 1]}%
   user{/Subtype/Link/A<</Type/Action/S/URI/URI(#1)>>}%
  \relax
 }%
 \providecommand\@@endlink[0]{\pdfendlink}%
}%
\providecommand \url  [0]{\begingroup\@sanitize \@url }%
\providecommand \@url [1]{\endgroup\@href {#1}{\urlprefix}}%
\providecommand \urlprefix [0]{URL }%
\providecommand \Eprint[0]{\href }%
\@ifxundefined \urlstyle {%
  \providecommand \doi [1]{doi:\discretionary{}{}{}#1}%
}{%
  \providecommand \doi [0]{doi:\discretionary{}{}{}\begingroup
  \urlstyle{rm}\Url }%
}%
\providecommand \doibase [0]{http://dx.doi.org/}%
\providecommand \Doi[1]{\href{\doibase#1}}%
\providecommand \bibAnnote [3]{%
  \BibitemShut{#1}%
  \begin{quotation}\noindent
    \textsc{Key:}\ #2\\\textsc{Annotation:}\ #3%
  \end{quotation}%
}%
\providecommand \bibAnnoteFile [2]{%
  \IfFileExists{#2}{\bibAnnote {#1} {#2} {\input{#2}}}{}%
}%
\providecommand \typeout [0]{\immediate \write \m@ne }%
\providecommand \selectlanguage [0]{\@gobble}%
\providecommand \bibinfo [0]{\@secondoftwo}%
\providecommand \bibfield [0]{\@secondoftwo}%
\providecommand \translation [1]{[#1]}%
\providecommand \BibitemOpen[0]{}%
\providecommand \bibitemStop [0]{}%
\providecommand \bibitemNoStop [0]{.\EOS\space}%
\providecommand \EOS [0]{\spacefactor3000\relax}%
\providecommand \BibitemShut [1]{\csname bibitem#1\endcsname}%
\bibitem{ARS70}%
  \BibitemOpen
  \bibfield{author}{%
  \bibinfo {author} {\bibfnamefont{A.~T.}\ \bibnamefont{Aldred}}, \bibinfo
  {author} {\bibfnamefont{B.~D.}\ \bibnamefont{Rainford}},\ and\ \bibinfo
  {author} {\bibfnamefont{M.~W.}\ \bibnamefont{Stringfellow}},\ }%
  \bibfield{journal}{%
  \bibinfo {journal} {Phys. Rev. Lett.}\ }%
  \textbf{\bibinfo {volume} {24}},\ \bibinfo {pages} {897} (\bibinfo {year}
  {1970})%
  \bibAnnoteFile{NoStop}{ARS70}%
\bibitem{HaZu72}%
  \BibitemOpen
  \bibfield{author}{%
  \bibinfo {author} {\bibfnamefont{R.}~\bibnamefont{Harris}}\ and\ \bibinfo
  {author} {\bibfnamefont{M.~J.}\ \bibnamefont{Zuckermann}},\ }%
  \bibfield{journal}{%
  \bibinfo {journal} {Phys. Rev. B}\ }%
  \textbf{\bibinfo {volume} {5}},\ \bibinfo {pages} {101} (\bibinfo {year}
  {1972})%
  \bibAnnoteFile{NoStop}{HaZu72}%
\bibitem{CrSc65}%
  \BibitemOpen
  \bibfield{author}{%
  \bibinfo {author} {\bibfnamefont{J.}~\bibnamefont{Crangle}}\ and\ \bibinfo
  {author} {\bibfnamefont{W.~R.}\ \bibnamefont{Scott}},\ }%
  \bibfield{journal}{%
  \bibinfo {journal} {J. Appl. Phys.}\ }%
  \textbf{\bibinfo {volume} {36}},\ \bibinfo {pages} {921} (\bibinfo {year}
  {1965})%
  \bibAnnoteFile{NoStop}{CrSc65}%
\bibitem{MTC74}%
  \BibitemOpen
  \bibfield{author}{%
  \bibinfo {author} {\bibfnamefont{A.~P.}\ \bibnamefont{Murani}}, \bibinfo
  {author} {\bibfnamefont{A.}~\bibnamefont{Tari}},\ and\ \bibinfo {author}
  {\bibfnamefont{B.~R.}\ \bibnamefont{Coles}},\ }%
  \bibfield{journal}{%
  \bibinfo {journal} {J. Phys. F: Met. Phys.}\ }%
  \textbf{\bibinfo {volume} {4}},\ \bibinfo {pages} {1769} (\bibinfo {year}
  {1974})%
  \bibAnnoteFile{NoStop}{MTC74}%
\bibitem{CKG81}%
  \BibitemOpen
  \bibfield{author}{%
  \bibinfo {author} {\bibfnamefont{T.~D.}\ \bibnamefont{Cheung}}, \bibinfo
  {author} {\bibfnamefont{J.~S.}\ \bibnamefont{Kouvel}},\ and\ \bibinfo
  {author} {\bibfnamefont{J.~W.}\ \bibnamefont{Garland}},\ }%
  \bibfield{journal}{%
  \bibinfo {journal} {Phys. Rev. B}\ }%
  \textbf{\bibinfo {volume} {23}},\ \bibinfo {pages} {1245} (\bibinfo {year}
  {1981})%
  \bibAnnoteFile{NoStop}{CKG81}%
\bibitem{BRLM83}%
  \BibitemOpen
  \bibfield{author}{%
  \bibinfo {author} {\bibfnamefont{S.~K.}\ \bibnamefont{Burke}}, \bibinfo
  {author} {\bibfnamefont{B.~D.}\ \bibnamefont{Rainford}}, \bibinfo {author}
  {\bibfnamefont{E.~J.}\ \bibnamefont{Lindley}},\ and\ \bibinfo {author}
  {\bibfnamefont{O.}~\bibnamefont{Moze}},\ }%
  \bibfield{journal}{%
  \bibinfo {journal} {J. Magn. Mag. Mat.}\ }%
  \textbf{\bibinfo {volume} {31{\bf-}34}},\ \bibinfo {pages} {545 } (\bibinfo
  {year} {1983})%
  \bibAnnoteFile{NoStop}{BRLM83}%
\bibitem{Vojt10}%
  \BibitemOpen
  \bibfield{author}{%
  \bibinfo {author} {\bibfnamefont{T.}~\bibnamefont{Vojta}},\ }%
  \bibfield{journal}{%
  \bibinfo {journal} {J. Low Temp. Phys.}\ }%
  \textbf{\bibinfo {volume} {161}},\ \bibinfo {pages} {299} (\bibinfo {year}
  {2010})%
  \bibAnnoteFile{NoStop}{Vojt10}%
\bibitem{NBKM99}%
  \BibitemOpen
  \bibfield{author}{%
  \bibinfo {author} {\bibfnamefont{M.}~\bibnamefont{Nicklas}}, \bibinfo
  {author} {\bibfnamefont{M.}~\bibnamefont{Brando}}, \bibinfo {author}
  {\bibfnamefont{G.}~\bibnamefont{Knebel}}, \bibinfo {author}
  {\bibfnamefont{F.}~\bibnamefont{Mayr}}, \bibinfo {author}
  {\bibfnamefont{W.}~\bibnamefont{Trinkl}},\ and\ \bibinfo {author}
  {\bibfnamefont{A.}~\bibnamefont{Loidl}},\ }%
  \bibfield{journal}{%
  \bibinfo {journal} {Phys. Rev. Lett.}\ }%
  \textbf{\bibinfo {volume} {82}},\ \bibinfo {pages} {4268} (\bibinfo {year}
  {1999})%
  \bibAnnoteFile{NoStop}{NBKM99}%
\bibitem{Lonz97}%
  \BibitemOpen
  \bibfield{author}{%
  \bibinfo {author} {\bibfnamefont{G.}~\bibnamefont{Lonzarich}},\ }%
  \enquote{\bibinfo {title} {{The Magnetic Electron}},}\ in\ \emph{\bibinfo
  {booktitle} {{Electron: A Centenary Volume}}},\ \bibinfo {editor} {edited by\
  \bibinfo {editor} {\bibfnamefont{M.}~\bibnamefont{Springford}}}\ (\bibinfo
  {publisher} {Cambridge University Press},\ \bibinfo {address} {Cambridge,
  U.K.},\ \bibinfo {year} {1997})\ Chap.~\bibinfo {chapter} {6}, p.\ \bibinfo
  {pages} {109}%
  \bibAnnoteFile{NoStop}{Lonz97}%
\bibitem{Stew01}%
  \BibitemOpen
  \bibfield{author}{%
  \bibinfo {author} {\bibfnamefont{G.~R.}\ \bibnamefont{Stewart}},\ }%
  \bibfield{journal}{%
  \bibinfo {journal} {Rev. Mod. Phys.}\ }%
  \textbf{\bibinfo {volume} {73}},\ \bibinfo {pages} {797} (\bibinfo {year}
  {2001})%
  \bibAnnoteFile{NoStop}{Stew01}%
\bibitem{HMW82}%
  \BibitemOpen
  \bibfield{author}{%
  \bibinfo {author} {\bibfnamefont{S.~C.}\ \bibnamefont{Ho}}, \bibinfo {author}
  {\bibfnamefont{I.}~\bibnamefont{Maartense}},\ and\ \bibinfo {author}
  {\bibfnamefont{G.}~\bibnamefont{Williams}},\ }%
  \bibfield{journal}{%
  \bibinfo {journal} {J. Appl. Phys.}\ }%
  \textbf{\bibinfo {volume} {53}},\ \bibinfo {pages} {2235} (\bibinfo {year}
  {1982})%
  \bibAnnoteFile{NoStop}{HMW82}%
\bibitem{Nieu75}%
  \BibitemOpen
  \bibfield{author}{%
  \bibinfo {author} {\bibfnamefont{G.}~\bibnamefont{Nieuwenhuys}},\ }%
  \bibfield{journal}{%
  \bibinfo {journal} {Adv. Phys.}\ }%
  \textbf{\bibinfo {volume} {24}},\ \bibinfo {pages} {515} (\bibinfo {year}
  {1975})%
  \bibAnnoteFile{NoStop}{Nieu75}%
\bibitem{PBKM84}%
  \BibitemOpen
  \bibfield{author}{%
  \bibinfo {author} {\bibfnamefont{R.~P.}\ \bibnamefont{Peters}}, \bibinfo
  {author} {\bibfnamefont{C.}~\bibnamefont{Buchal}}, \bibinfo {author}
  {\bibfnamefont{M.}~\bibnamefont{Kubota}}, \bibinfo {author}
  {\bibfnamefont{R.~M.}\ \bibnamefont{Mueller}},\ and\ \bibinfo {author}
  {\bibfnamefont{F.}~\bibnamefont{Pobell}},\ }%
  \bibfield{journal}{%
  \bibinfo {journal} {Phys. Rev. Lett.}\ }%
  \textbf{\bibinfo {volume} {53}},\ \bibinfo {pages} {1108} (\bibinfo {year}
  {1984})%
  \bibAnnoteFile{NoStop}{PBKM84}%
\bibitem{KSP95}%
  \BibitemOpen
  \bibfield{author}{%
  \bibinfo {author} {\bibfnamefont{Y.}~\bibnamefont{Kondo}}, \bibinfo {author}
  {\bibfnamefont{K.}~\bibnamefont{Swieca}},\ and\ \bibinfo {author}
  {\bibfnamefont{F.}~\bibnamefont{Pobell}},\ }%
  \bibfield{journal}{%
  \bibinfo {journal} {J. Low Temp. Phys.}\ }%
  \textbf{\bibinfo {volume} {100}},\ \bibinfo {pages} {195} (\bibinfo {year}
  {1995})%
  \bibAnnoteFile{NoStop}{KSP95}%
\bibitem{RCKS94}%
  \BibitemOpen
  \bibfield{author}{%
  \bibinfo {author} {\bibfnamefont{B.}~\bibnamefont{Rainford}}, \bibinfo
  {author} {\bibfnamefont{R.}~\bibnamefont{Cywinski}}, \bibinfo {author}
  {\bibfnamefont{S.}~\bibnamefont{Kilcoyne}},\ and\ \bibinfo {author}
  {\bibfnamefont{C.}~\bibnamefont{Scott}},\ }%
  \bibfield{journal}{%
  \bibinfo {journal} {Hyperfine Interact.}\ }%
  \textbf{\bibinfo {volume} {85}},\ \bibinfo {pages} {323} (\bibinfo {year}
  {1994})%
  \bibAnnoteFile{NoStop}{RCKS94}%
\bibitem{HHWH92}%
  \BibitemOpen
  \bibfield{author}{%
  \bibinfo {author} {\bibfnamefont{O.}~\bibnamefont{Hartmann}}, \bibinfo
  {author} {\bibfnamefont{S.~W.}\ \bibnamefont{Harris}}, \bibinfo {author}
  {\bibfnamefont{R.}~\bibnamefont{W\"appling}},\ and\ \bibinfo {author}
  {\bibfnamefont{R.}~\bibnamefont{Hempelmann}},\ }%
  \bibfield{journal}{%
  \bibinfo {journal} {Phys. Scripta}\ }%
  \textbf{\bibinfo {volume} {45}},\ \bibinfo {pages} {402} (\bibinfo {year}
  {1992})%
  \bibAnnoteFile{NoStop}{HHWH92}%
\bibitem{Karl95}%
  \BibitemOpen
  \bibfield{author}{%
  \bibinfo {author} {\bibfnamefont{E.~B.}\ \bibnamefont{Karlsson}},\ }%
  \emph{\bibinfo {title} {{Solid State Phenomena As Seen by Muons, Protons, and
  Excited Nuclei}}}\ (\bibinfo {publisher} {Clarendon Press},\ \bibinfo
  {address} {Oxford},\ \bibinfo {year} {1995})%
  \bibAnnoteFile{NoStop}{Karl95}%
\bibitem{Noak99}%
  \BibitemOpen
  \bibfield{author}{%
  \bibinfo {author} {\bibfnamefont{D.~R.}\ \bibnamefont{Noakes}},\ }%
  \bibfield{journal}{%
  \bibinfo {journal} {J. Phys.:\ Condens. Matter}\ }%
  \textbf{\bibinfo {volume} {11}},\ \bibinfo {pages} {1589} (\bibinfo {year}
  {1999})%
  \bibAnnoteFile{NoStop}{Noak99}%
\bibitem{Sche85}%
  \BibitemOpen
  \bibfield{author}{%
  \bibinfo {author} {\bibfnamefont{A.}~\bibnamefont{Schenck}},\ }%
  \emph{\bibinfo {title} {{Muon Spin Rotation Spectroscopy: Principles and
  Applications in Solid State Physics}}}\ (\bibinfo {publisher} {A. Hilger},\
  \bibinfo {address} {Bristol \& Boston},\ \bibinfo {year} {1985})%
  \bibAnnoteFile{NoStop}{Sche85}%
\bibitem{LKC99}%
  \BibitemOpen
  \bibinfo {editor} {\bibfnamefont{S.~L.}\ \bibnamefont{Lee}}, \bibinfo
  {editor} {\bibfnamefont{S.~H.}\ \bibnamefont{Kilcoyne}},\ and\ \bibinfo
  {editor} {\bibfnamefont{R.}~\bibnamefont{Cywinski}},\ eds.,\ \emph{\bibinfo
  {title} {Muon Science: Muons in Physics, Chemistry and Materials}},\ \bibinfo
  {series} {Scottish Universities Summer School in Physics}\ No.~\bibinfo
  {number} {51}\ (\bibinfo {publisher} {Institute of Physics Publishing},\
  \bibinfo {address} {Bristol \& Philadelphia},\ \bibinfo {year} {1999})%
  \bibAnnoteFile{NoStop}{LKC99}%
\bibitem{KNH01}%
  \BibitemOpen
  \bibfield{author}{%
  \bibinfo {author} {\bibfnamefont{G.~M.}\ \bibnamefont{Kalvius}}, \bibinfo
  {author} {\bibfnamefont{D.~R.}\ \bibnamefont{Noakes}},\ and\ \bibinfo
  {author} {\bibfnamefont{O.}~\bibnamefont{Hartmann}},\ }%
  in\ \emph{\bibinfo {booktitle} {Handbook on the Physics and Chemistry of Rare
  Earths}},\ Vol.~\bibinfo {volume} {32},\ \bibinfo {editor} {edited by\
  \bibinfo {editor} {\bibfnamefont{K.~A.}\ \bibnamefont{Gschneidner},
  \bibfnamefont{Jr.}}, \bibinfo {editor}
  {\bibfnamefont{L.}~\bibnamefont{Eyring}},\ and\ \bibinfo {editor}
  {\bibfnamefont{G.~H.}\ \bibnamefont{Lander}}}\ (\bibinfo {publisher}
  {Elsevier},\ \bibinfo {year} {2001})\ Chap.\ \bibinfo {chapter} {206}, pp.\
  \bibinfo {pages} {55--451}%
  \bibAnnoteFile{NoStop}{KNH01}%
\bibitem{YaDdR11}%
  \BibitemOpen
  \bibfield{author}{%
  \bibinfo {author} {\bibfnamefont{A.}~\bibnamefont{Yaouanc}}\ and\ \bibinfo
  {author} {\bibfnamefont{P.}~\bibnamefont{Dalmas~de R\'eotier}},\ }%
  \emph{\bibinfo {title} {{Muon Spin Rotation, Relaxation, and Resonance:
  Applications to Condensed Matter}}},\ International series of monographs on
  physics\ (\bibinfo {publisher} {Oxford University Press},\ \bibinfo {address}
  {New York},\ \bibinfo {year} {2011})%
  \bibAnnoteFile{NoStop}{YaDdR11}%
\bibitem{Note1}%
  \BibitemOpen
  \bibinfo {note} {The main magnetic field is applied parallel to the muon
  beam, since deviation of the beam is prohibitive for perpendicular fields
  greater than a few tens of millitesla. For TF-$\mu $SR in the main field, the
  separator fields are increased in order to rotate the muon spin $90^\circ
  $.}%
  \bibAnnoteFile{Stop}{Note1}%
\bibitem{Note2}%
  \BibitemOpen
  \bibinfo {note} {See, e.g., Ref.~\protect ~\cite {YaDdR11}, Chap.~5}%
  \bibAnnoteFile{NoStop}{Note2}%
\bibitem{WKW90}%
  \BibitemOpen
  \bibfield{author}{%
  \bibinfo {author} {\bibfnamefont{Z.}~\bibnamefont{Wang}}, \bibinfo {author}
  {\bibfnamefont{H.~P.}\ \bibnamefont{Kunkel}},\ and\ \bibinfo {author}
  {\bibfnamefont{G.}~\bibnamefont{Williams}},\ }%
  \bibfield{journal}{%
  \bibinfo {journal} {J. Phys.:\ Condens. Matter}\ }%
  \textbf{\bibinfo {volume} {2}},\ \bibinfo {pages} {4173} (\bibinfo {year}
  {1990})%
  \bibAnnoteFile{NoStop}{WKW90}%
\bibitem{Note3}%
  \BibitemOpen
  \bibinfo {note} {Muon diffusion would result in ``motionally-narrowed''
  relaxation characterized by a single exponential decay.}%
  \bibAnnoteFile{Stop}{Note3}%
\bibitem{KuTo67}%
  \BibitemOpen
  \bibfield{author}{%
  \bibinfo {author} {\bibfnamefont{R.}~\bibnamefont{Kubo}}\ and\ \bibinfo
  {author} {\bibfnamefont{T.}~\bibnamefont{Toyabe}},\ }%
  in\ \emph{\bibinfo {booktitle} {{Magnetic Resonance and Relaxation}}},\
  \bibinfo {editor} {edited by\ \bibinfo {editor}
  {\bibfnamefont{R.}~\bibnamefont{Blinc}}}\ (\bibinfo {publisher}
  {North-Holland},\ \bibinfo {address} {Amsterdam},\ \bibinfo {year} {1967})\
  pp.\ \bibinfo {pages} {810--823}%
  \bibAnnoteFile{NoStop}{KuTo67}%
\bibitem{HUIN79}%
  \BibitemOpen
  \bibfield{author}{%
  \bibinfo {author} {\bibfnamefont{R.~S.}\ \bibnamefont{Hayano}}, \bibinfo
  {author} {\bibfnamefont{Y.~J.}\ \bibnamefont{Uemura}}, \bibinfo {author}
  {\bibfnamefont{J.}~\bibnamefont{Imazato}}, \bibinfo {author}
  {\bibfnamefont{N.}~\bibnamefont{Nishida}}, \bibinfo {author}
  {\bibfnamefont{T.}~\bibnamefont{Yamazaki}},\ and\ \bibinfo {author}
  {\bibfnamefont{R.}~\bibnamefont{Kubo}},\ }%
  \bibfield{journal}{%
  \bibinfo {journal} {Phys. Rev. B}\ }%
  \textbf{\bibinfo {volume} {20}},\ \bibinfo {pages} {850} (\bibinfo {year}
  {1979})%
  \bibAnnoteFile{NoStop}{HUIN79}%
\bibitem{Sche85a}%
  \BibitemOpen
  \bibinfo {note} {See, e.g., Ref.~\protect~\cite{Sche85}, Sec.~4.4.}%
  \bibAnnoteFile{Stop}{Sche85a}%
\bibitem{Kubo81}%
  \BibitemOpen
  \bibfield{author}{%
  \bibinfo {author} {\bibfnamefont{R.}~\bibnamefont{Kubo}},\ }%
  \bibfield{journal}{%
  \bibinfo {journal} {Hyperfine Interact.}\ }%
  \textbf{\bibinfo {volume} {8}},\ \bibinfo {pages} {731} (\bibinfo {year}
  {1981})%
  \bibAnnoteFile{NoStop}{Kubo81}%
\bibitem{Uemu99}%
  \BibitemOpen
  \bibinfo {note} {Y.~J. Uemura, in Ref.~\protect~\cite{LKC99}, p.~85.}%
  \bibAnnoteFile{Stop}{Uemu99}%
\bibitem{WaWa74}%
  \BibitemOpen
  \bibfield{author}{%
  \bibinfo {author} {\bibfnamefont{R.~E.}\ \bibnamefont{Walstedt}}\ and\
  \bibinfo {author} {\bibfnamefont{L.~R.}\ \bibnamefont{Walker}},\ }%
  \bibfield{journal}{%
  \bibinfo {journal} {Phys. Rev. B}\ }%
  \textbf{\bibinfo {volume} {9}},\ \bibinfo {pages} {4857} (\bibinfo {year}
  {1974})%
  \bibAnnoteFile{NoStop}{WaWa74}%
\bibitem{Darb67}%
  \BibitemOpen
  \bibfield{author}{%
  \bibinfo {author} {\bibfnamefont{M.~I.}\ \bibnamefont{Darby}},\ }%
  \bibfield{journal}{%
  \bibinfo {journal} {Br. J. Appl. Phys.}\ }%
  \textbf{\bibinfo {volume} {18}},\ \bibinfo {pages} {1415} (\bibinfo {year}
  {1967})%
  \bibAnnoteFile{NoStop}{Darb67}%
\bibitem{Prat07}%
  \BibitemOpen
  \bibfield{author}{%
  \bibinfo {author} {\bibfnamefont{F.~L.}\ \bibnamefont{Pratt}},\ }%
  \bibfield{journal}{%
  \bibinfo {journal} {J. Phys.:\ Condens. Matter}\ }%
  \textbf{\bibinfo {volume} {19}},\ \bibinfo {pages} {456207} (\bibinfo {year}
  {2007})%
  \bibAnnoteFile{NoStop}{Prat07}%
\bibitem{UYHS85}%
  \BibitemOpen
  \bibfield{author}{%
  \bibinfo {author} {\bibfnamefont{Y.~J.}\ \bibnamefont{Uemura}}, \bibinfo
  {author} {\bibfnamefont{T.}~\bibnamefont{Yamazaki}}, \bibinfo {author}
  {\bibfnamefont{D.~R.}\ \bibnamefont{Harshman}}, \bibinfo {author}
  {\bibfnamefont{M.}~\bibnamefont{Senba}},\ and\ \bibinfo {author}
  {\bibfnamefont{E.~J.}\ \bibnamefont{Ansaldo}},\ }%
  \bibfield{journal}{%
  \bibinfo {journal} {Phys. Rev. B}\ }%
  \textbf{\bibinfo {volume} {31}},\ \bibinfo {pages} {546} (\bibinfo {year}
  {1985})%
  \bibAnnoteFile{NoStop}{UYHS85}%
\bibitem{Note4}%
  \BibitemOpen
  \bibinfo {note} {By ``Ni spins'' we mean the atomic spins together with any
  associated Pd spin polarization.}%
  \bibAnnoteFile{Stop}{Note4}%
\bibitem{TaCo71}%
  \BibitemOpen
  \bibfield{author}{%
  \bibinfo {author} {\bibfnamefont{A.}~\bibnamefont{Tari}}\ and\ \bibinfo
  {author} {\bibfnamefont{B.~R.}\ \bibnamefont{Coles}},\ }%
  \bibfield{journal}{%
  \bibinfo {journal} {J. Phys. F: Met. Phys.}\ }%
  \textbf{\bibinfo {volume} {1}},\ \bibinfo {pages} {L69} (\bibinfo {year}
  {1971})%
  \bibAnnoteFile{NoStop}{TaCo71}%
\bibitem{BeTo76}%
  \BibitemOpen
  \bibfield{author}{%
  \bibinfo {author} {\bibfnamefont{J.}~\bibnamefont{Beille}}\ and\ \bibinfo
  {author} {\bibfnamefont{R.}~\bibnamefont{Tournier}},\ }%
  \bibfield{journal}{%
  \bibinfo {journal} {J. Phys. F: Met. Phys.}\ }%
  \textbf{\bibinfo {volume} {6}},\ \bibinfo {pages} {621} (\bibinfo {year}
  {1976})%
  \bibAnnoteFile{NoStop}{BeTo76}%
\bibitem{Chou76}%
  \BibitemOpen
  \bibfield{author}{%
  \bibinfo {author} {\bibfnamefont{G.}~\bibnamefont{Chouteau}},\ }%
  \bibfield{journal}{%
  \bibinfo {journal} {Physica B+C}\ }%
  \textbf{\bibinfo {volume} {84}},\ \bibinfo {pages} {25 } (\bibinfo {year}
  {1976})%
  \bibAnnoteFile{NoStop}{Chou76}%
\bibitem{SaKo78}%
  \BibitemOpen
  \bibfield{author}{%
  \bibinfo {author} {\bibfnamefont{D.}~\bibnamefont{Sain}}\ and\ \bibinfo
  {author} {\bibfnamefont{J.~S.}\ \bibnamefont{Kouvel}},\ }%
  \bibfield{journal}{%
  \bibinfo {journal} {Phys. Rev. B}\ }%
  \textbf{\bibinfo {volume} {17}},\ \bibinfo {pages} {2257} (\bibinfo {year}
  {1978})%
  \bibAnnoteFile{NoStop}{SaKo78}%
\bibitem{KiGa91}%
  \BibitemOpen
  \bibfield{author}{%
  \bibinfo {author} {\bibfnamefont{N.}~\bibnamefont{Kioussis}}\ and\ \bibinfo
  {author} {\bibfnamefont{J.~W.}\ \bibnamefont{Garland}},\ }%
  \bibfield{journal}{%
  \bibinfo {journal} {Phys. Rev. Lett.}\ }%
  \textbf{\bibinfo {volume} {67}},\ \bibinfo {pages} {366} (\bibinfo {year}
  {1991})%
  \bibAnnoteFile{NoStop}{KiGa91}%
\bibitem{BCLR82}%
  \BibitemOpen
  \bibfield{author}{%
  \bibinfo {author} {\bibfnamefont{S.~K.}\ \bibnamefont{Burke}}, \bibinfo
  {author} {\bibfnamefont{R.}~\bibnamefont{Cywinski}}, \bibinfo {author}
  {\bibfnamefont{E.~J.}\ \bibnamefont{Lindley}},\ and\ \bibinfo {author}
  {\bibfnamefont{B.~D.}\ \bibnamefont{Rainford}},\ }%
  \bibfield{journal}{%
  \bibinfo {journal} {J. Appl. Phys.}\ }%
  \textbf{\bibinfo {volume} {53}},\ \bibinfo {pages} {8079} (\bibinfo {year}
  {1982})%
  \bibAnnoteFile{NoStop}{BCLR82}%
\bibitem{Pfle12pc}%
  \BibitemOpen
  \bibinfo {note} {C. Pfleiderer (private communication)}%
  \bibAnnoteFile{NoStop}{Pfle12pc}%
\bibitem{PBFK10}%
  \BibitemOpen
  \bibfield{author}{%
  \bibinfo {author} {\bibfnamefont{C.}~\bibnamefont{Pfleiderer}}, \bibinfo
  {author} {\bibfnamefont{P.}~\bibnamefont{B\"oni}}, \bibinfo {author}
  {\bibfnamefont{C.}~\bibnamefont{Franz}}, \bibinfo {author}
  {\bibfnamefont{T.}~\bibnamefont{Keller}}, \bibinfo {author}
  {\bibfnamefont{A.}~\bibnamefont{Neubauer}}, \bibinfo {author}
  {\bibfnamefont{P.}~\bibnamefont{Niklowitz}}, \bibinfo {author}
  {\bibfnamefont{P.}~\bibnamefont{Schmakat}}, \bibinfo {author}
  {\bibfnamefont{M.}~\bibnamefont{Schulz}}, \bibinfo {author}
  {\bibfnamefont{Y.-K.}\ \bibnamefont{Huang}}, \bibinfo {author}
  {\bibfnamefont{J.}~\bibnamefont{Mydosh}}, \bibinfo {author}
  {\bibfnamefont{M.}~\bibnamefont{Voj\-ta}}, \bibinfo {author}
  {\bibfnamefont{W.}~\bibnamefont{Duncan}}, \bibinfo {author}
  {\bibfnamefont{F.}~\bibnamefont{Grosche}}, \bibinfo {author}
  {\bibfnamefont{M.}~\bibnamefont{Brando}}, \bibinfo {author}
  {\bibfnamefont{M.}~\bibnamefont{Deppe}}, \bibinfo {author}
  {\bibfnamefont{C.}~\bibnamefont{Geibel}}, \bibinfo {author}
  {\bibfnamefont{F.}~\bibnamefont{Steglich}}, \bibinfo {author}
  {\bibfnamefont{A.}~\bibnamefont{Krimmel}},\ and\ \bibinfo {author}
  {\bibfnamefont{A.}~\bibnamefont{Loidl}},\ }%
  \bibfield{journal}{%
  \bibinfo {journal} {J. Low Temp. Phys.}\ }%
  \textbf{\bibinfo {volume} {161}},\ \bibinfo {pages} {167} (\bibinfo {year}
  {2010})%
  \bibAnnoteFile{NoStop}{PBFK10}%
\bibitem{Odod83}%
  \BibitemOpen
  \bibfield{author}{%
  \bibinfo {author} {\bibfnamefont{J.~C.}\ \bibnamefont{Ododo}},\ }%
  \bibfield{journal}{%
  \bibinfo {journal} {J. Phys. F: Met. Phys.}\ }%
  \textbf{\bibinfo {volume} {13}},\ \bibinfo {pages} {1291} (\bibinfo {year}
  {1983})%
  \bibAnnoteFile{NoStop}{Odod83}%
\bibitem{StWi00}%
  \BibitemOpen
  \bibfield{author}{%
  \bibinfo {author} {\bibfnamefont{P.~A.}\ \bibnamefont{Stampe}}\ and\ \bibinfo
  {author} {\bibfnamefont{G.}~\bibnamefont{Williams}},\ }%
  \bibfield{journal}{%
  \bibinfo {journal} {Solid State Commun.}\ }%
  \textbf{\bibinfo {volume} {113}},\ \bibinfo {pages} {607 } (\bibinfo {year}
  {2000})%
  \bibAnnoteFile{NoStop}{StWi00}%
\bibitem{EGMG12}%
  \BibitemOpen
  \bibfield{author}{%
  \bibinfo {author} {\bibfnamefont{N.}~\bibnamefont{Egetenmeyer}}, \bibinfo
  {author} {\bibfnamefont{J.~L.}\ \bibnamefont{Gavilano}}, \bibinfo {author}
  {\bibfnamefont{A.}~\bibnamefont{Maisuradze}}, \bibinfo {author}
  {\bibfnamefont{S.}~\bibnamefont{Gerber}}, \bibinfo {author}
  {\bibfnamefont{D.~E.}\ \bibnamefont{MacLaughlin}}, \bibinfo {author}
  {\bibfnamefont{G.}~\bibnamefont{Seyfarth}}, \bibinfo {author}
  {\bibfnamefont{D.}~\bibnamefont{Andreica}}, \bibinfo {author}
  {\bibfnamefont{A.}~\bibnamefont{Desilets-Benoit}}, \bibinfo {author}
  {\bibfnamefont{A.~D.}\ \bibnamefont{Bianchi}}, \bibinfo {author}
  {\bibfnamefont{C.}~\bibnamefont{Baines}}, \bibinfo {author}
  {\bibfnamefont{R.}~\bibnamefont{Khasanov}}, \bibinfo {author}
  {\bibfnamefont{Z.}~\bibnamefont{Fisk}},\ and\ \bibinfo {author}
  {\bibfnamefont{M.}~\bibnamefont{Kenzelmann}},\ }%
  \bibfield{journal}{%
  \bibinfo {journal} {Phys. Rev. Lett.}\ }%
  \textbf{\bibinfo {volume} {108}},\ \bibinfo {pages} {177204} (\bibinfo {year}
  {2012})%
  \bibAnnoteFile{NoStop}{EGMG12}%
\bibitem{Grif69}%
  \BibitemOpen
  \bibfield{author}{%
  \bibinfo {author} {\bibfnamefont{R.~B.}\ \bibnamefont{Griffiths}},\ }%
  \bibfield{journal}{%
  \Doi{10.1103/PhysRevLett.23.17}{\bibinfo {journal} {Phys. Rev. Lett.}}\ }%
  \textbf{\bibinfo {volume} {23}},\ \bibinfo {pages} {17} (\bibinfo {year}
  {1969})%
  \bibAnnoteFile{NoStop}{Grif69}%
\end{thebibliography}

%

\end{document}